\documentclass[12pt,a4paper]{article}
\usepackage{graphicx}
\usepackage{dcolumn}
\usepackage{bm}
\usepackage{amsmath,amsfonts,amssymb,mathrsfs}
\usepackage{slashed, MnSymbol}
\usepackage{braket,xcolor}
\usepackage{verbatim}
\usepackage{subcaption}
\usepackage{multirow}
\usepackage{amsfonts,mathtools,resizegather}
\usepackage[normalem]{ulem}
\usepackage[utf8]{inputenc}
\usepackage{multicol}
\usepackage{subcaption}
\usepackage[normalem]{ulem}
\usepackage{caption,jheppub}

\usepackage[symbol]{footmisc}

\footnotetext{Authors' names are listed in alphabetical order.}

\title{\boldmath {On Krylov complexity in open systems: an approach via bi-Lanczos algorithm}
 }
\author[a,b]{Aranya Bhattacharya,}
\author[c,d]{Pratik Nandy,}
\author[a]{Pingal Pratyush Nath,}
\author[a,e]{and Himanshu Sahu}

\definecolor{himcolor}{HTML}{52B788}

\affiliation[a]{Centre for High Energy Physics, Indian Institute of Science,\\ C.V. Raman Avenue, Bangalore 560012, India.}
\affiliation[b]{Institute of Physics, Jagiellonian University, Łojasiewicza 11, 30-348 Kraków, Poland}
\affiliation[c]{Center for Gravitational Physics and Quantum Information,\\ Yukawa Institute for Theoretical Physics, Kyoto University,\\
Kitashirakawa Oiwakecho, Sakyo-ku, Kyoto 606-8502, Japan}
\affiliation[d]{RIKEN Interdisciplinary Theoretical and Mathematical Sciences Program (iTHEMS),\\
Wako, Saitama 351-0198, Japan}
\affiliation[e]{Dept. of Instrumentation \& Applied Physics, Indian Institute of Science,\\ C.V. Raman Avenue, Bangalore 560012, India}

\emailAdd{aranyabhattacharya13@gmail.com}
\emailAdd{pratik@yukawa.kyoto-u.ac.jp}
\emailAdd{pingalnath@iisc.ac.in}
\emailAdd{himanshusah1@iisc.ac.in}

\abstract{Continuing the previous initiatives \cite{Bhattacharya:2022gbz, Bhattacharjee:2022lzy}, we pursue the exploration of operator growth and Krylov complexity in dissipative open quantum systems. In this paper, we resort to the bi-Lanczos algorithm generating two bi-orthogonal Krylov spaces, which individually generate non-orthogonal subspaces. Unlike the previously studied Arnoldi iteration, this algorithm renders the Lindbladian into a purely tridiagonal form, thus opening up a possibility to study a wide class of dissipative integrable and non-integrable systems by computing Krylov complexity at late times. Our study relies on two specific systems, the dissipative transverse-field Ising model (TFIM) and the dissipative interacting XXZ chain. We find that, for the weak coupling, initial Lanczos coefficients can efficiently distinguish integrable and chaotic evolution before the dissipative effect sets in, which results in more fluctuations in higher Lanczos coefficients. This results in the equal saturation of late-time complexity for both integrable and chaotic cases, making the notion of late-time chaos dubious.}

\begin{document}
\maketitle
\flushbottom

\section{Introduction}
Krylov complexity (K-complexity in short) has recently been proven to be a useful probe in diagnosing scrambling and quantum chaos \cite{Parker:2018yvk, Barbon:2019wsy}. In particular, Krylov-complexity growth is conjectured to upper bound the growth of Lyapunov exponent at any finite temperature \cite{Parker:2018yvk, Avdoshkin:2019trj}. Correspondingly, the Lanczos coefficients, an output of the Lanczos algorithm \cite{Lanczos1950AnIM, viswanath1994recursion}, which is the basic machinery of computing K-complexity, show distinct ``asymptotic'' behavior in its integrable and chaotic evolution, albeit some exceptions \cite{Dymarsky:2021bjq, Bhattacharjee:2022vlt}. The behavior of Lanczos coefficients and correspondingly K-complexity captures both the early time as well as late time chaos \cite{Rabinovici:2020ryf, Rabinovici:2021qqt, Rabinovici:2022beu, Hornedal:2022pkc}. While the initial coefficients usually show different scalings for integrable and chaotic cases characterizing the early time distinction, in late times, the two cases are distinguished by observing the fluctuations in the coefficients, dubbed as the Krylov localization \cite{Rabinovici:2021qqt}. This phenomenon is manifested in the late-time saturation of K-complexity, with chaotic systems showing higher saturation than integrable counterparts.

In our previous works \cite{Bhattacharya:2022gbz, Bhattacharjee:2022lzy}, we initiated a systematic study of such formalism in open quantum systems (see \cite{Liu:2022god, Bhattacharjee:2023uwx, Kawabata:2022cpr, Zhang:2022knu, Kawabata:2022osw, Schuster:2022bot, Weinstein:2022yce, Matsoukas-Roubeas:2022odk, Garcia-Garcia:2021elz, Garcia-Garcia:2022xsh, Jia:2022reh} for some recent works on operator growth and chaos on non-Hermitian and open quantum systems). A long-term motivation for such a study is to sharpen the understanding of black hole physics by considering open quantum field theories \cite{Loganayagam:2022zmq}. Unlike the closed systems, the open system operator evolution (under the realm of Markovian dynamics \cite{Breuer2007}) is characterized by a non-unitary evolution through the exponentiated non-Hermitian Lindbladian $\mathcal{L}_o \, \cdot=[H,\,\cdot\,]+i T$, where the second term represents the non-Hermitian part \cite{Lindblad1976, Gorini}. The knowledge of the system and its interaction with the environment is assumed through the Hamiltonian and some Lindblad (jump) operators. In such cases, the previous study of spin chains \cite{Bhattacharya:2022gbz} and dissipative Sachdev-Ye-Kitaev (SYK) \cite{Bhattacharjee:2022lzy} generalize the result of \cite{Parker:2018yvk}, by proposing two sets of Lanczos coefficients, characterizing the operator growth in generic systems. Both use Arnoldi iteration \cite{Arnoldi1951ThePO} which generalizes the Lanczos algorithm to the non-Hermitian evolution, the coefficients show distinct and consistent exploration of the Krylov basis for integrable and chaotic parameter regimes. However, on the Arnoldi basis, the Lindbladian does not reduce to a pure tridiagonal matrix, rather it features an upper-Hessenberg form. Due to such a finitely large number of coefficients (however small they are), the computation of the complexity becomes increasingly difficult with system sizes. However, some analytic results have been obtained in dissipative SYK \cite{Bhattacharjee:2022lzy}, where the complexity appears to be suppressed and inversely proportional to the dissipation parameter for a particular class of Lindbladian \cite{Kulkarni:2021gtt}. Although the results feature the dynamics of an inherently chaotic system, a substantial class of dissipative integrable systems remains largely unexplored.

In this paper, we bridge this gap by performing systematic studies of spin chains that can be smoothly tuned to integrable and chaotic regimes. We also focus on a specific integrable interacting system. The first example is the transverse-field Ising model (TFIM) while the second one is the interacting XXZ chain. We work with a different algorithm known as the bi-Lanczos algorithm which is an efficient algorithm for evolution under a non-Hermitian Lindbladian superoperator. This is conceptually different from the previously studied Arnoldi iteration \cite{Bhattacharya:2022gbz, Bhattacharjee:2022lzy}. Both algorithm essentially tackles the same problem yet falls into a distinct class of projection method. The bi-Lanczos algorithm deals with the evolving “bra” and “ket” vectors separately, therefore constructing two Krylov bases instead of one. We find that a successful implementation of the bi-Lanczos algorithm explores the Krylov basis efficiently. Unlike the Arnoldi iteration, in the bi-orthogonal basis (we continue to call it Krylov basis unless any confusion arises), the Lindbladian can be recast in a fully tridiagonal form. The imaginary diagonal coefficients contain the information on the dissipation, while the off-diagonal elements depend on the integrable nature of the Hamiltonian. This simple tridiagonal form enables us to compute the complexity efficiently, especially its late-time saturation. However, the effect of dissipation makes the late-time chaos more involved compared to the closed system. In particular, we carefully study the role of dissipation in the suppression of complexity which is ascribed as ``Krylov localization'' \cite{Rabinovici:2021qqt}. In both of our models, featuring integrable and chaotic regimes, we find similar results.

The rest of the paper is structured as follows. In section \ref{krylovalgo}, we explain the closed system Lanczos algorithm and the bi-Lanczos algorithm applied to the open system evolution. We also define the normalized version of the K-complexity that we will be using throughout. In section \ref{models}, we introduce our main model (TFIM) and the results, especially subsection \ref{results} present our main results for open TFIM.  Finally, in section \ref{conc}, we conclude with some future directions. Various appendices contain supporting information. In Appendix \ref{appa}, we show how the bi-Lanczos algorithm reduces to Lanczos for unitary evolution. In Appendix \ref{appb}, we discuss the open XXZ model and the results obtained there, which agree with the results of open TFIM, presented in the main text. In Appendix \ref{biL_spread}, we discuss some extensions of the bi-Lanczos algorithm to spread complexity.

\section{Krylov algorithms for closed and open systems} \label{krylovalgo}

In this section, we briefly introduce the construction of the Krylov basis in both closed and open systems. In the closed system, the straightforward procedure is to apply the Lanczos algorithm. However, in the open systems, there exist two different algorithms, Arnoldi iteration, and the bi-Lanczos algorithm. Both algorithms come with their own advantages and disadvantages. Arnoldi iteration was previously applied in dissipative spin chains and dissipative SYK \cite{Bhattacharya:2022gbz, Bhattacharjee:2022lzy}.\footnote{Closed system Lanczos basis has also been applied to open systems \cite{Liu:2022god}. Refer to \cite{Bhattacharjee:2022lzy} for a comprehensive discussion between different approaches.} Here, we introduce the bi-Lanczos algorithm in subsection \ref{biLsec}, which will be our main concern in this paper.

\subsection{Closed system and Lanczos algorithm}

We start our discussion with the closed systems where the Liouvillian is Hermitian. Hence, the operator evolution is unitary. In this case, the time-evolved operator in Heisenberg evolution can be written as
\begin{equation*}
    O(t)=e^{iHt}\,O(0)\,e^{-iHt}=O(0) + it[H,O(0)] + \frac{(it)^2}{2}[H,[H,O(0)]] + \cdots = e^{i\mathcal{L}t}O(0)\,,
\end{equation*}
where $\mathcal{L} \,\, \cdot = [H, \,\cdot\,]$ is the Liouvillian. A convenient way to study the growth of a simple operator is to realize them as states, namely $O(0) \equiv |O_0)$, and to introduce a notion of an inner product. We will be dealing with the infinite-temperature inner product, also known as the Frobenius norm 
\begin{equation}
    (O_m|O_n)= \frac{\mathrm{Tr}\, [O_m^{\dagger}O_n]}{\mathrm{Tr}\,[\,\mathbb{I}\,]}\,.
\end{equation}
Our task is to build a fully orthonormal basis out of the repetitive action of $\mathcal{L}$ on $|O_0)$. The conventional way to do this is to perform the well-known Lanczos algorithm. The resulting basis, known as the Krylov basis satisfies 
\begin{equation}
    \mathcal{L}|O_n) = b_n|O_{n-1}) - b_{n+1}|O_{n+1})\,.
\end{equation}
with the matrix element of the Liouvillian given by a tridiagonal matrix
\begin{equation}
    \mathcal{L}_{m n}  =(O_m|\mathcal{L}|O_n) ~~~~ \Rightarrow ~~~~ \mathcal{L} =\begin{pmatrix} 0&b_{1}&&&&0\\b_{1}&  0& b_{2}&&&\\&b_{2}&0&\ddots &&\\&&\ddots &\ddots &b_{m}&\\&&&b_{m}&0&
    \ddots\\0&&&&\ddots&\ddots\\\end{pmatrix}\,.
\end{equation}
In this basis, the diagonal elements vanish and the full information of the growth of the operator contains in the set of coefficients $b_n$, known as the Lanczos coefficients. In finite dimensional systems, the set of $b_n$ is finite and terminated once the Krylov space is exhausted. Nevertheless, an asymptotic growth can still be understood by considering a suitable region of $b_n$, usually termed as ``Lanczos ascent'' \cite{Kar:2021nbm}. For chaotic systems, the growth is conjectured to be linear in $n$,  \cite{Parker:2018yvk}, while the integrable systems usually show sublinear growth \cite{Parker:2018yvk} (some exceptions were also found in \cite{Dymarsky:2021bjq, Bhattacharjee:2022vlt}).

A consistent understanding of the spread of an operator in the Krylov basis can be developed by expanding the time-evolved operator in the following manner
\begin{equation}
    |O (t)) = \sum_{n=0}^{\mathcal{K}-1} i^n \phi_n(t) |O_n)\,,
\end{equation}
where the time-dependent coefficients, known as Krylov wavefunctions $\phi_n(t)$ are understood as the probability amplitudes of finding the operators in the $n$-th Krylov basis at time $t$. For a unitary evolution, this results in the probability conservation equation, namely $\sum_{n=0}^{\mathcal{K}-1}|\phi_n(t)|^2=1$. The Krylov complexity (K-complexity in short) in this basis is then defined as the average position of the Krylov operator satisfying $\hat{K} |O_n) =  n \, |O_n)$:
\begin{equation}
    K(t) = (O(t) | \hat{K} | O(t)) = \sum_{n=0}^{\mathcal{K}-1} n \, |\phi_n(t)|^2\,.
\end{equation}
The growth of Lanczos coefficients directly reflects in the growth of K-complexity. The linear growth of Lanczos coefficients gives rise to the exponential growth of complexity, while the sublinear growth of Lanczos coefficients correspond to the polynomial growth of K-complexity \cite{Parker:2018yvk, Barbon:2019wsy}.


\subsection{Open system: Arnoldi and bi-Lanczos algorithms}\label{biLsec}
In the case of open systems, the analog of Liouvillian within the Born-Markovian approximation is the Lindbladian ($\mathcal{L}_o$), which is non-Hermitian. This makes the evolution $e^{i\mathcal{L}_o t}$ becomes non-unitary. In this case, apart from the usual commutator term, there exists an extra piece constructed out of the Lindblad (jump) operators. These Lindblad operators, built out of system operators, reflect the interaction of the system with its environment.

In such cases, the usual Lanczos algorithm fails to work efficiently, precisely due to its non-Hermiticity. Generalizing to the open systems leads to consideration of the Arnoldi iteration that can generate a systematic orthonormal basis from the non-Hermitian Lindbladian. However, as found in \cite{Bhattacharya:2022gbz, Bhattacharjee:2022lzy}, it recasts the Lindbladian into an upper-Hessenberg form, which is not ideally suitable to compute probabilities and complexity due to the existence of a substantial number of elements in the matrix. In this work, we use the bi-Lanczos algorithm, a complementary approach compared to the Arnoldi iteration. Without the environmental interaction, it reduces to the usual Lanczos algorithm. This algorithm, however, turns out to be efficient in computing probabilities and complexity since it recasts the Lindbladian into a pure tridiagonal form. The primary idea is to generate two separate Krylov basis sets; in particular, one defines them as two sets of bi-orthonormal vectors $|p_n\rrangle$ and $|q_n\rrangle$ such that
\begin{equation}\label{biortho}
    \llangle q_m|p_n \rrangle =\delta_{mn}\,,
\end{equation}
where the ``double braces'' indicates that the vectors are constructed using the Choi-Jamio\l{}kowski (CJ) isomorphism \cite{CHOI1975285, JAMIOLKOWSKI1972275}. This efficiently takes care of the non-Hermitian vectors after the action of the non-Hermitian Lindbladian. The reason is that a general element $|p_n \rrangle$ of the ket-Krylov basis will be non-Hermitian and will generically not be orthonormal to itself i.e., $\llangle p_m|p_n \rrangle \neq \delta_{mn}$. However, once two separate bases are generated consistently as two bi-orthogonal vector spaces, the construction makes sure that Eq.\eqref{biortho} is satisfied for all the elements of the two bases. This notion was first hinted in \cite{Bhattacharya:2022gbz, Bhattacharjee:2022lzy, parkerthesis}\footnote{We thank Xiangyu Cao for suggesting this to us during the preparation of the work \cite{Bhattacharya:2022gbz}.}. In this case, the action of $\mathcal{L}_o$ on the ket space $|p_n\rrangle$ from left is realized the generalized Lanczos algorithm while the action of $\mathcal{L}_o$ on bra space $\llangle q_n|$ is realized by acting $\mathcal{L}_o^{\dagger}$ on the ket space generated by $|q_n \rrangle$. The starting vector is usually taken as $|p_0\rrangle=|q_{0}\rrangle$ before applying the Lindbladian. However, in general $|p_n\rrangle \neq |q_n\rrangle$ for $n > 0$. This extends the Lanczos algorithm to the non-Hermitian case by a two-sided iterative algorithm, namely by bi-orthogonalizing via a two-sided Gram-Schmidt procedure. The two Krylov sequences \cite{biL1}
\begin{align*}
   \mathbb{K}^j(\mathcal{L}_{o},p_0)&= \{p_0, \mathcal{L}_{o} \, p_0, \mathcal{L}_{o}^2 \,p_0, \ldots\}\,,   \\
    \mathbb{K}^j(\mathcal{L}_{o}^\dagger,q_0)&= \{q_0, \mathcal{L}_{o}^\dagger \, q_0,(\mathcal{L}_{o}^\dagger)^2 \,q_0, \ldots \}\,,
\end{align*}
themselves do not generate orthonormal subspace. However, the two sequences of vectors $\{q_i\}$ and $\{p_i\}$ are generated using the three-term recurrences
\begin{align}
    c_{j+1} |p_{j+1}\rrangle &=\mathcal{L}_{o} |p_j\rrangle -a_j |p_j \rrangle  -b_{j} |p_{j-1}\rrangle\, \label{bilanczosbasic} \\
    b^*_{j+1} |q_{j+1}\rrangle &= \mathcal{L}_{o}^\dagger |q_j\rrangle - a^*_j |q_j\rrangle -c^*_{j} |q_{j-1}\rrangle\,. \label{bilanczosbasic2}
\end{align}
The vectors $\{p_i\}$ and $\{q_i\}$ are called \textit{bi-Lanczos} vectors, they span $ \mathbb{K}^j(\mathcal{L}_{o},p_0)$ and $\mathbb{K}^j(\mathcal{L}_{o}^\dagger,q_0)$ respectively and are \emph{bi-orthonormal}. 

In order to express the Lindbladian in this bi-orthonormal basis, it is customary to define vectors in matrix notation as follows 
\begin{align}
    P = (p_0~\, p_1~ \, \ldots ~ \, p_m\, \ldots) \,, ~~~~~~  Q =(q_0~\, q_1 ~\, \ldots ~\, q_m\, \ldots)\,.
\end{align}
Then the Lindbladian is expressed as
\begin{align}
    [\mathcal{L}_{o}] = Q^\dagger \mathcal{L}_{o} P  = \begin{pmatrix} a_{0}&b_{1}&&&&0\\c_{1}&  a_{1}& b_{2}&&&\\&c_{2}&a_{2}&\ddots &&\\&&\ddots &\ddots &b_{m}&\\&&&c_{m}&a_{m}&
    \ddots\\0&&&&\ddots&\ddots\\\end{pmatrix}\,. \label{ld}
\end{align}
Below, we outline the detailed algorithm to construct such bases. The algorithm goes as follows \cite{gaaf2017infinite, bbiL}:

\begin{enumerate}
    \item Let $|p_0\rrangle, |q_0\rrangle \in \mathbb{C}^n$ be arbitrary vectors with $\llangle q_0|p_0\rrangle = 1$., i.e., we choose $|p_0\rrangle = |q_0\rrangle \equiv |O_0)$ at the initial step.
    \item The initial iteration steps are given as follows:
    \begin{enumerate}
        \item Let $|r'_0\rrangle =\mathcal{L}_{o} |p_0\rrangle$ and $|s'_0\rrangle =\mathcal{L}_{o}^\dagger  |q_0\rrangle$.
        \item Compute the inner product $a_0 = \llangle q_0|r'_0 \rrangle$.
        \item Define $|r_0\rrangle =|r_0'\rrangle -a_0 |p_0\rrangle$ and $|s_0\rrangle = |s'_0\rrangle -a^*_0 |q_0\rrangle$.
    \end{enumerate}
    
    \item for $j=1,2, \ldots$, perform the following steps: 
    \begin{enumerate}
        \item Compute the inner product $\omega_{j}=\llangle r_{j-1}|s_{j-1} \rrangle$.
        \item Compute the norm $c_{j}= \sqrt{|\omega_{j}|}$ and $b_{j} = \omega^*_{j}/c_{j}$.
        \item If $c_{j}\neq0$, let 
        \begin{align}
            |p_j\rrangle = \frac{|r_{j-1}\rrangle}{c_{j}}~~~~ \& ~~~~ |q_j\rrangle =\frac{|s_{j-1}\rrangle}{b^*_{j}}\,.
        \end{align}
        \item If required, perform the full orthogonalization\footnote{We have explicitly performed this full re-orthogonalization after each iteration and therefore the basis is exactly bi-orthonormal.}
        \begin{align*}
            |p_j\rrangle = |p_j\rrangle - \sum_{i=0}^{j-1} \llangle q_i|p_j \rrangle\, |p_i \rrangle\,, ~~~~~~
            |q_j\rrangle = |q_j\rrangle - \sum_{i=0}^{j-1} \llangle p_i|q_j \rrangle \, |q_i \rrangle\,.
        \end{align*}
        \item Let $|r'_j \rrangle = \mathcal{L}_{o} |p_j \rrangle $  and $|s'_j\rrangle = \mathcal{L}_{o}^\dagger  |q_j\rrangle$. 
        \item Compute $a_{j} = \llangle q_j|r'_j \rrangle$.
        \item Define the vectors:
         \begin{align*}
            |r_j\rrangle = |r'_j \rrangle  -a_{j}|p_j\rrangle -b_{j} |p_{j-1}\rrangle\,, ~~
            |s_j\rrangle = |s_j'\rrangle -a^*_{j} |q_j\rrangle -c^*_{j}|q_{j-1}\rrangle\,. 
        \end{align*}
        and go back to step $3$.
            \end{enumerate}
    \item If $c_j=0$ for some $j = \mathcal{K}-1$, where $\mathcal{K}$ is the Krylov dimension, let $P= (p_0~\,p_1 ~\,\ldots~\,p_{\mathcal{K}-1})$, and $Q= (q_0~\,q_1 ~\,\ldots~\,q_{\mathcal{K}-1})$. The Lindbladian is then given by $[\mathcal{L}_o]=Q^{\dagger} \mathcal{L}_o P$.
\end{enumerate}

\subsection{Properties based on observation}
Here we list down the observations based on the results derived from the application of the bi-Lanczos algorithm on various models. As we have stated earlier, the bi-Lanczos algorithm gives the Lindbladian of the form \eqref{ld}. The matrix elements have the following properties:
\begin{enumerate}
    \item $\mathrm{Im}(c_n)=0$ since $c_n = \sqrt{|\omega_{n}|}$ which is always real.
    \item In general, $b_n \neq c_n$. To see this, assume that $b_n=c_n$, so that by definition 
        \begin{align*}
        c_n &= \sqrt{|\omega_{n}|} ~~~~ \& ~~~~ b_n =\frac{\omega^*_{n}}{c_n}\,, \\
        b_n&= c_n ~~\Rightarrow~~ |\omega_{n}| = \omega_{n}\,.
    \end{align*}
    which in turn means that $\omega_{n}$ must be positive real number but from definition $\omega_{n}=\llangle r_{n-1}|s_{n-1} \rrangle$ which in general is a complex number. 
    \item However, $|b_n|=|c_n|$ since by construction 
    \begin{align*}
        c_n &= \sqrt{|\omega_{n}|} ~~\Rightarrow~~ |\omega_{n}|=c^2_n\,.\\
        b_n &=\frac{\omega^*_{n}}{c_n}~~\Rightarrow~~ |b_n|= \frac{|\omega_{n}|}{|c_n|} = |c_n|\,.
    \end{align*}
\end{enumerate}
Note that in this case, by construction, there are no further matrix elements. Since $|b_n|=|c_n|$, they can differ only by a phase factor. In fact, we numerically find that $b_n=c_n\in \mathbb{R}^{+}$ and diagonals $a_n$ are purely imaginary ($a_n=i a_n$) and positive\footnote{
In some special cases, some of the purely imaginary diagonals might be negative, although the trace of the tridiagonalized matrix will still be positive. We thank Niklas H\"ornedal for pointing this out to us.} ($a_n\in \mathbb{C}^{+}$). Therefore in further calculations, we take $b_n=c_n=|b_n|$ and $a_n=i |a_n|$. We, therefore, end up with an effective tridiagonal matrix of the following form
\begin{equation}
    \mathcal{L}_o= \begin{pmatrix}i|a_{0}|& |b _{1}|&&&&0\\ |b_1| & i|a _{1}|&|b _{2}|&&&\\& |b_{2}|& i|a _{2}| &\ddots &&\\&&\ddots &\ddots & |b_{m}|&\\&&& |b _{m}|& i|a _{m}|& \ddots \\0&&&& \ddots & \ddots \\\end{pmatrix}\,. \label{LI}
\end{equation}
In later sections, we consider specific examples and construct the corresponding Lindbladian.

\subsection{Probability and Krylov complexity}

Unlike isolated closed systems, in open quantum systems, one would in general expect the probability of finding the operator within the system Krylov basis to decay with time. This is expected because the operator spread in this case is not limited to the system subspace and can always get support from the environment. Within the open system treatment, although we construct a systematic bi-orthogonal basis, we find the above-mentioned expectation results from the non-Hermitian Lindbladian. More precisely, this happens due to the purely imaginary diagonals in the tridiagonal representation of the Lindbladian after implementing the bi-Lanczos algorithm. In the most general setting, the bi-Lanczos algorithm leads the following expansions for the evolution of the state (ket) and its dual (bra) vectors
\begin{equation}
    \ket{O(t)} =\sum_{i=0}^{\mathcal{K}-1} i^n \phi_n (t) |p_n \rrangle \,,~~~~ \bra{O(t)} = \sum_{i=0}^{\mathcal{K}-1} \llangle q_n| (-i)^n \psi^{*}_n(t)\,.
\end{equation}
Using the above expression, the probability of finding the operator at $n$-th element of the bi-orthogonal Krylov basis becomes the following
\begin{equation}
    P(t) =\braket{O(t)|O(t)}=\sum_{n=0}^{\mathcal{K}-1} \psi^{*}_n(t) \phi_n(t)\,.
\end{equation}
However, since in the cases studied in this paper, we find $b_n=c_n\in \mathbb{R}^{+}$, if we look at equations \eqref{bilanczosbasic} and \eqref{bilanczosbasic2}, which would give the corresponding equations\footnote{The equation with $|q_{j+1}\rangle$ would give wavefunction $\psi_n$ because there the $\psi_n^{*}$ is associated to the ``bra'' version of the $q_n$ vectors.} for the wavefunctions $\phi_n$ and $\psi_n$, it is easy to see that taking the complex conjugate of \eqref{bilanczosbasic2}, one would find the equation followed by $\psi_n^{*}$. It turns out that this equation is exactly the same as the equation followed by the complex conjugate of $\phi_n$. In other words, if $b_n=c_n$, one can write $\psi_n^{*}=(\phi_n)^{*}$.

Given the above fact, in our studies, we can effectively treat the probability still as $P(t)=\sum_{n=0}^{\mathcal{K}-1} \phi^{*}_n(t) \phi_n(t)$. Note that the probability for a closed system with unitary dynamics will always be equal to one as the bi-Lanczos algorithm reduces to the Lanczos algorithm in that case (see Appendix \ref{appa}). However, in the case of the open systems, the evolution is non-unitary, and hence the probability is expected to decay due to the dissipative effects. This decay is expected to be very similar to the decay due to decoherence discussed in the equation (5.9) of \cite{Matsoukas-Roubeas:2022odk}. Similarly, we can try to (naively) define the  K-complexity in the bi-Lanczos basis as 
\begin{equation}
    K(t)= \sum_{n=0}^{\mathcal{K}-1} n \,\phi^{*}_n(t) \phi_n(t)\,.
\end{equation}
This usual notion of complexity is problematic because the probability itself decays with time. Therefore, for the open quantum systems, we study the normalized K-complexity\footnote{From now onwards, we refer to the normalized K-complexity simply as K-complexity.}
\begin{equation}\label{bikrylovform}
    K_o (t)= \frac{\sum_{n=0}^{\mathcal{K}-1} n\, \phi^{*}_n(t) \phi_n(t)}{\sum_{n=0}^{\mathcal{K}-1} \phi^{*}_n(t) \phi_n(t)}\,.
\end{equation}
The corresponding differential equation followed by these coefficients is the following
\begin{equation}
    \dot{\phi}_n(t) = c_n \phi_{n-1}(t)-b_{n+1}\phi_{n+1}(t)+i a_n \phi_n (t)\,,
\end{equation}
and an equivalent one for the $\psi_n(t)$. In the RHS of the discretized differential equation, the coefficient multiplied by $\phi_n(t)$ is $i a_n$. However, as mentioned previously, the diagonals themselves are purely imaginary (by replacing $a_n=i|a_n|\rightarrow i a_n=-|a_n|$ and $c_n=b_n=|b_n|$ ignoring the overall phase factors in $c_n$ as finite size errors). Hence the differential equation becomes
\begin{equation}
    \dot{\phi}_n(t) = |b_n| \phi_{n-1}(t)-|b_{n+1}|\phi_{n+1}(t)-|a_n| \phi_n (t)\,,
\end{equation}
where the particular term proportional to $\phi_n(t)$ in the RHS now makes each of the solved $\phi_n(t)$ go through an exponentially decaying behavior. This precisely results in the decay of probability as well. 

For a finite number of Lanczos coefficients, We solve $\mathcal{K}$ such differential equations to get the behavior of probability and the complexity valid till very late times. To do so, we solve the simplified matrix differential equation as follows
\begin{equation}
   \frac{d\Phi(t)}{dt}=\frac{d}{dt} \begin{pmatrix}
\phi_0(t)\\
\phi_1(t) \\
\phi_2(t)\\
\vdots\\
\phi_{\mathcal{K}-2}(t)\\
\phi_{\mathcal{K}-1}(t)\\
\end{pmatrix}=
\begin{pmatrix}-|a_{0}|& -|b _{1}|&&&&0\\ |b_1| & -|a _{1}|& -|b _{2}|&&&\\& |b_{2}|& -|a _{2}| &\ddots &&\\&&\ddots &\ddots & -|b_{\mathcal{K}-2}|&\\&&& |b _{\mathcal{K}-2}|& -|a _{\mathcal{K}-2}|& -|b _{\mathcal{K}-1}|\\0&&&& |b_{\mathcal{K}-1}| & -|a _{\mathcal{K}-1}|\\\end{pmatrix} \begin{pmatrix}
\phi_0(t)\\
\phi_1(t) \\
\phi_2(t)\\
\vdots\\
\phi_{\mathcal{K}-2}(t)\\
\phi_{\mathcal{K}-1}(t)\\
\end{pmatrix}=\mathcal{S}\cdot\Phi(t) \,.
\end{equation}
Notice that the big matrix $\mathcal{S} \equiv \mathcal{S}_{\mathcal{K} \times \mathcal{K}}$ in the RHS is slightly different from the tridiagonal representation of the non-Hermitian Lindbladian $\mathcal{L}_o$ that we obtained earlier. This matrix is written by looking at the form of the equation followed by the $\phi_n(t)$. Now solving the matrix differential equation for the column vector $\Phi(t)$, we can afterward get the probability and un-normalized K-complexity by simply computing the norm of the column vectors $\Phi(t)$ ($P(t)=\text{Norm}[\Phi(t)]^2$) and  $\sqrt{\mathscr{K}(t)}_n=\sqrt{n} \phi_n(t)$ respectively. Remember that once we solve the column vector $\Phi(t)$, we have the $\phi_n(t)$ for all $n$. Therefore, it is easy to form the column vector $\sqrt{\mathscr{K}(t)}$
\begin{equation}
    \sqrt{\mathscr{K}(t)}=\begin{pmatrix}
0\,\phi_0(t)\\
\sqrt{1}\,\phi_1(t) \\
\sqrt{2}\,\phi_2(t)\\
\vdots\\
\sqrt{\mathcal{K}-2}\,\phi_{\mathcal{K}-2}(t)\\
\sqrt{\mathcal{K}-1}\,\phi_{\mathcal{K}-1}(t)\\
\end{pmatrix}\,,
\end{equation}
the norm of which is the un-normalized K-complexity, given by $K(t)=\sum_{n=0}^{\mathcal{K}-1} n|\phi_n(t)|^2=\text{Norm}[\sqrt{\mathscr{K}(t)}]^2$.

Finally, we get the normalized K-complexity for the open systems by computing the following
\begin{equation}
    K_o(t)=\frac{\text{Norm}[\sqrt{\mathscr{K}(t)}]^2}{\text{Norm}[\Phi(t)]^2}\,. \label{nKr}
\end{equation}
Note that this definition is essentially the same as Eq. \eqref{bikrylovform}, with the advantage of being numerically efficient in our case. In the next section, we introduce our model and discuss the numerical results.

\section{Setup and results: Transverse-Field Ising Model}\label{models}

In open systems, the Born-Markov approximation gives the quantum master equation, which dictates the non-unitary evolution of the open system \cite{Lindblad1976, Gorini}
\begin{equation}
    	\mathcal{L}_o [\,\bullet\,] = [H,\,\bullet\,]-i\sum_{k}\big[L_k^{\dagger} \bullet L_k-\frac{1}{2}\{L_k^{\dagger}L_k,\bullet\}\big]\,,
\end{equation}
where $\mathcal{L}_o$ is the Lindbladian superoperator acting on an operator, denoted by the ``bullet'' $\bullet$. Unlike isolated closed systems, the evolution is in terms of a non-Hermitian Lindbladian 
\begin{equation}
    O(t)=e^{i\mathcal{L}_o t} \,O(0)\,.
\end{equation}
The non-Hermitian part of the Lindbladian is constructed out of the Lindblad operators $L_k$, which are, in fact, signatures of the interaction of the system with its environment. Therefore, it is the interaction that generates and introduces non-Hermiticity in the Lindbladian. See \cite{Bhattacharya:2022gbz, Bhattacharjee:2022lzy} for more details. In the following, we concentrate on the model we study, which is the open transverse-field Ising model (TFIM). We discuss the results for Lanczos coefficients and complexity found in the integrable and non-integrable regimes. Similar results can be found for the open interacting XXZ spin chain which we present in Appendix \ref{appb}. 

\subsection{Hamiltonian: TFIM}

The transverse-field Ising model (TFIM) Hamiltonian describes the behavior of a collection of interacting spins arranged in a lattice, with each spin described by Pauli matrices. The Hamiltonian includes two terms: the first represents the interactions between the spins, and the second represents the interactions between the spins and an external transverse field.  The behavior of the system is controlled by the ratio of the couplings $g$ and $h$, with different regimes of behavior observed as this ratio varied. The TFIM is of great interest due to its close relationship with the quantum phase transition phenomena and the ability to use it to investigate the properties of quantum systems. The Hamiltonian is given by
\begin{align}
    H_{\mathrm{TFIM}} = - \sum_{j=1}^{N-1}  \sigma^{z}_{j} \sigma^{z}_{j+1} - g \sum_{j=1}^{N} \sigma^{x}_j - h \sum_{j=1}^{N} \sigma^{z}_{j}\,, \label{tfim}
\end{align}
where $g$ and $h$ are the coupling parameters. Given $h=0$, the Hamiltonian is integrable for all values of $g$, where it can be mapped to the free-fermionic model \cite{sachdev_2011}. On the other hand, it goes away from integrability for nonzero longitudinal coupling $h$. In the integrable regime, we choose  $g=1, h=0$, while we choose $g=-1.05, h=0.5$ for the non-integrable/chaotic regime \cite{PhysRevLett.106.050405}.\footnote{In all further results reported on the TFIM model, we will refer to these values of $g$ and $h$ as integrable and non-integrable/chaotic limits.}

We encode the interaction between the system and the environment by the following jump operators \cite{Bhattacharya:2022gbz, PhysRevLett.123.254101}
\begin{align}
	L_{-1} &= \sqrt{\alpha}\,\sigma_1^{+}\,,~~~~\, L_0=\sqrt{\alpha}\,\sigma_1^{-}\,, \nonumber \\
	L_{N+1} & = \sqrt{\alpha}\,\sigma_N^{+}\,, ~~~~L_{N+2} = \sqrt{\alpha}\,\sigma_N^{-}\,, 
	~~~~L_i =\sqrt{\gamma} \,\sigma_{i}^z\,, ~~~~ i =1, 2, \cdots, N\,.\label{jump}
\end{align}
where $\sigma_j^{\pm}= (\sigma_j^{x}\pm i \sigma_j^y)/2$. The set of operators $L_k$ with $k = -1,\, 0,\, N+1,\, \text{and}\, N+2$ captures the boundary amplitude damping with amplitude $\alpha > 0$, and the bulk dephasing is encoded by the operators $L_i, ~i =1, 2, \cdots, N$ with amplitude $\gamma > 0$. 

\subsection{Results} \label{results}

Here we implement the bi-Lanczos algorithm for the open TFIM as explained previously. We get an effective tridiagonal matrix for the Lindbladian. The $\alpha = \gamma = 0$ case corresponds to the closed TFIM case, for which we get back the closed system behavior. For all our examples, we choose the system size $N = 6$. 

\begin{figure}[t]
   \centering
\begin{subfigure}[b]{0.46\textwidth}
\centering
\includegraphics[width=\textwidth]{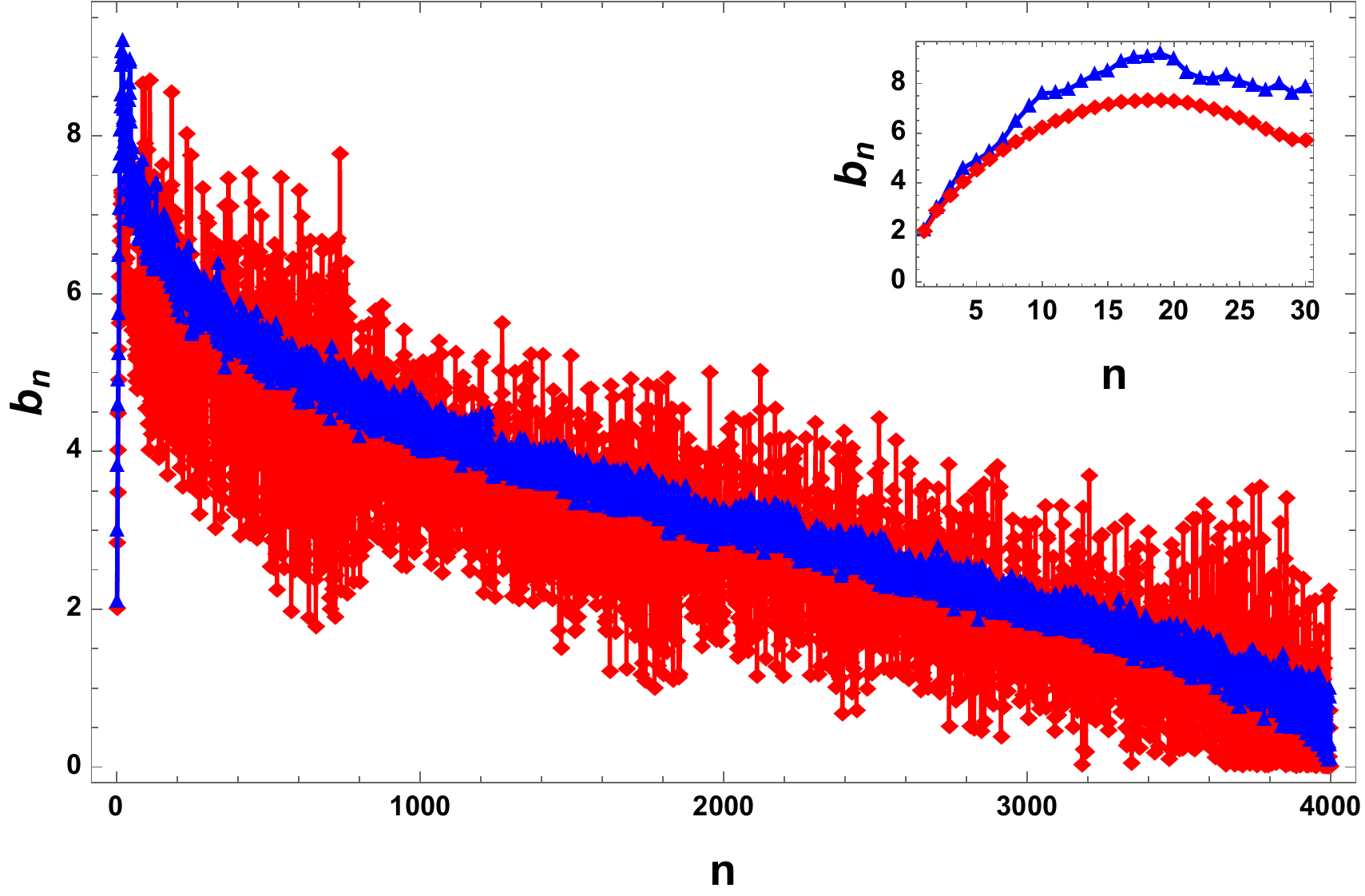}
\caption{}
\end{subfigure}
\hfill
\begin{subfigure}[b]{0.46\textwidth}
\centering
\includegraphics[width=\textwidth]{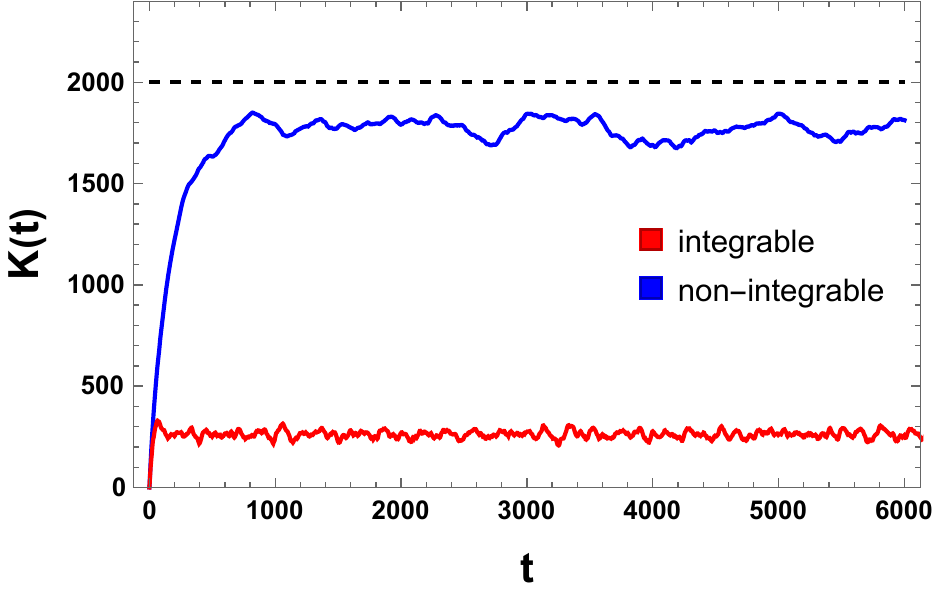}
\caption{}
\end{subfigure}
\caption{(a) Growth of Lanczos coefficients $b_n$ for the probe operator $\sigma_3^z$ in integrable ($g=1$, $h=0$, red) and chaotic ($g=-1.05$, $h=0.5$, blue) limits as mentioned below Eq. \eqref{tfim} by implementing the bi-Lanczos algorithm. The system size is $N = 6$. Since this is a closed system, the Lanczos and bi-Lanczos algorithm yields the same results. The behavior of the first few coefficients is shown in the inset image. In the Lanczos descent, there are more fluctuations for the integrable case compared to the non-integrable one. (b) The behavior of K-complexity for integrable (red) and chaotic (blue) limits. Observe that the saturation value in the chaotic limit is higher ($\sim \mathcal{K}/2$, dashed black line) than the saturation value in the integrable limit.} \label{fig:Lanclosed}
\end{figure}
	
\subsubsection{Closed systems} 
We are interested in studying the operator growth for the initial operator at site $i=3$ (in the system size $N=6$), i.e., $O_3 \equiv \sigma_3^z$. The Krylov dimension, in this case, is of the order $\mathcal{K}\sim 4000$. For the closed case, we get an exact agreement with the results of \cite{Rabinovici:2022beu} (see also \cite{PhysRevE.107.024217}).

\begin{itemize}
    \item The diagonal coefficients $a_n$ are vanishing. Also, the other two sets of coefficients, $b_n$ and $c_n$, which are in general of the same magnitude, become exactly equal.

    \item The integrable and the chaotic $b_n$ for small $n$ (up to $30$) are differentiable (inset of Fig.\,\ref{fig:Lanclosed} (a)). In the integrable case, the initial coefficients grow sub-linearly, whereas, in the non-integrable case, they grow almost linearly.
    
    \item For large $n$ iterations, the integrable coefficients exhibit more fluctuations than the chaotic ones. See Fig.\,\ref{fig:Lanclosed} (a). This results in a lower saturation of K-complexity for the integrable regime, compared to the chaotic ones (Fig.\,\ref{fig:Lanclosed} (b)). This phenomenon, in integrable case, is usually attributed to the presence of stronger disorder in the Lanczos sequence, known as the Krylov localization \cite{Rabinovici:2021qqt}. However, the saturation value of complexity for the chaotic case is $\sim K/2\sim 2000$, which is in agreement with \cite{Rabinovici:2022beu}.
\end{itemize}

\subsubsection{Open systems}
For the open systems, the extra component is the diagonal elements in the tridiagonal representation of the Lindbladian as explained in the previous section. The purely imaginary nature of the diagonals results in a decay of the amplitude associated with each of the $\phi_n(t)$. The behavior of the diagonals ($a_n$) and off diagonals ($b_n$) are as follows.

\begin{figure}[t]
\centering
\begin{subfigure}[b]{0.32\textwidth}
\centering
\includegraphics[width=\textwidth]{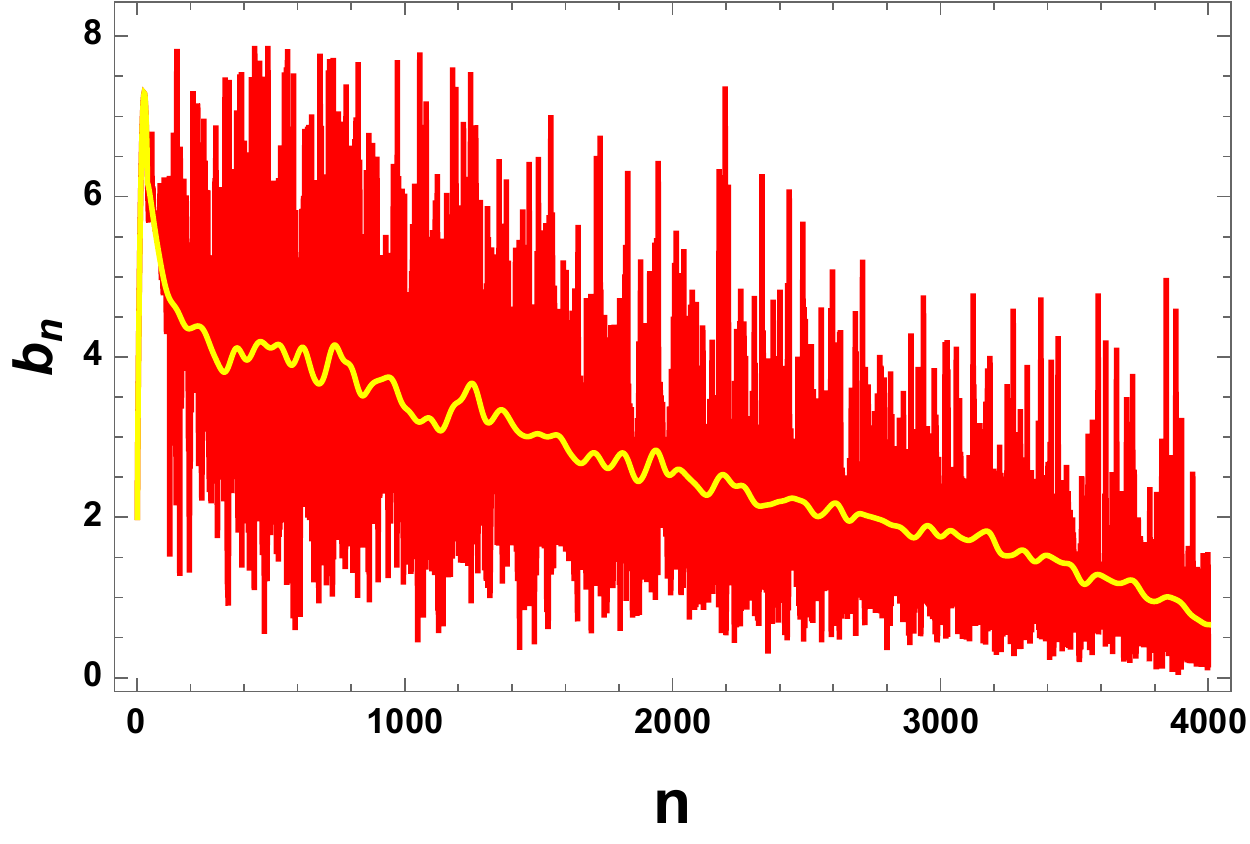}
\caption{}
\end{subfigure}
\hfill
\begin{subfigure}[b]{0.32\textwidth}
\centering
\includegraphics[width=\textwidth]{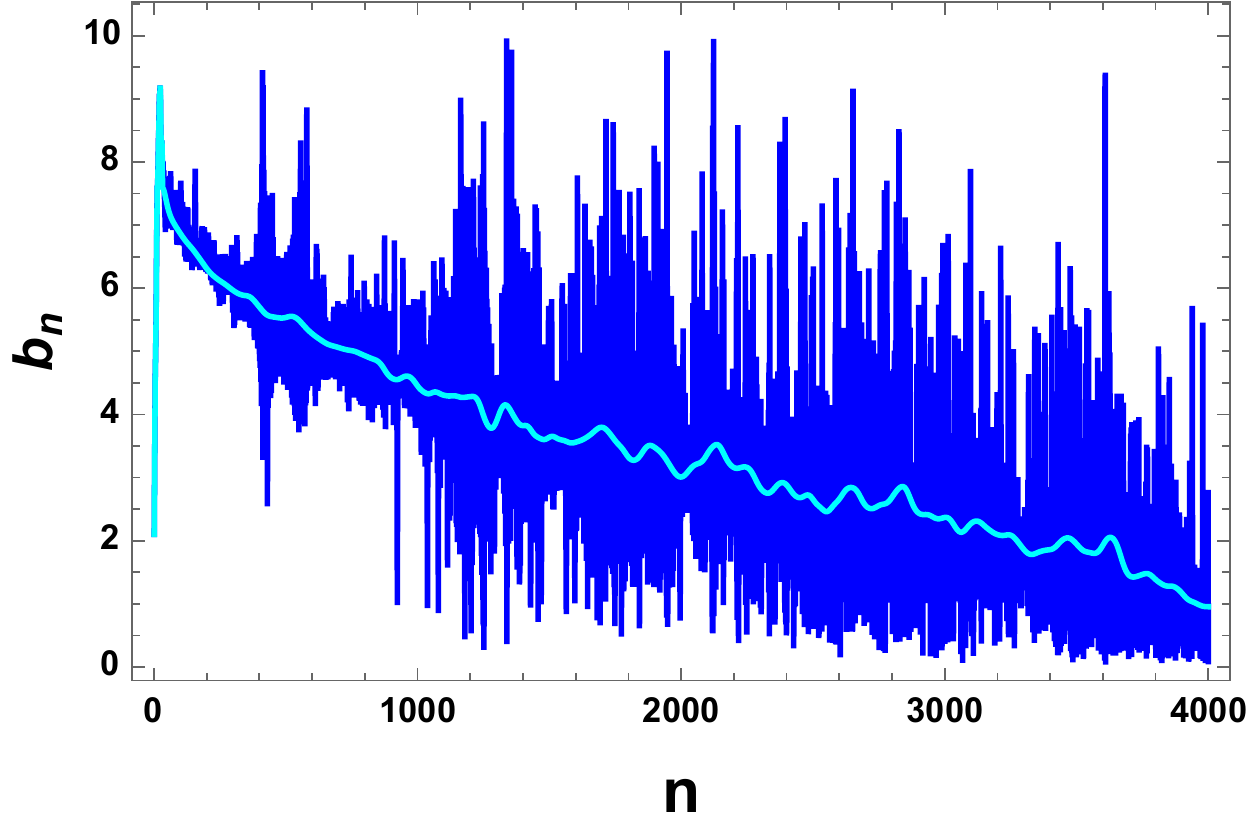}
\caption{}
\end{subfigure}
\hfill
\begin{subfigure}[b]{0.32\textwidth}
\centering
\includegraphics[width=\textwidth]{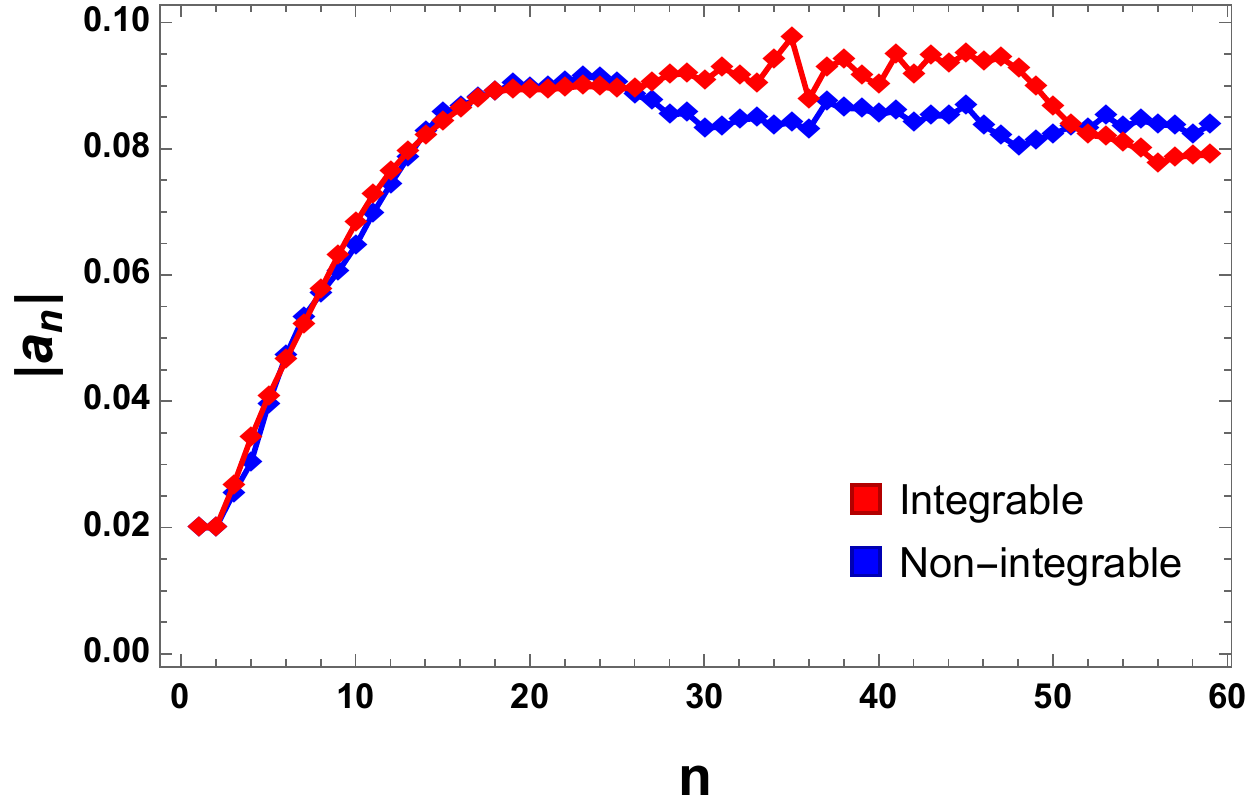}
\caption{}
\end{subfigure}
\caption{ Growth of Lanczos coefficients for (a) integrable ($g=1$, $h=0$) and (b) chaotic ($g=-1.05$, $h=0.5$) systems for $\alpha=0.01, \, \gamma=0.01$. The system size is $N = 6$. The yellow and cyan lines in the respective plots show the averaged values, done only to better understand the overall Lanczos descent. One can easily see that in the case of open systems, even the chaotic $b_n$ have fluctuations for large $n$ resulting in the same saturation value of K-complexity as the integrable case. The information of integrability, therefore, washes out at higher times, or higher $n$. (c) shows the comparison between $|a_n|$ plots for integrable and non-integrable cases with $\alpha=0.01$, $\gamma=0.01$.} \label{fig:bnopenTFIM}
\end{figure}

\begin{table}
\begin{center}
\begin{subtable}[b]{0.46\textwidth}
\centering
\begin{tabular}{||c c c||} 
 \hline
 $\gamma$ & $\eta_{\mathrm{int}}$ & $\eta_{\mathrm{non-int}}$  \\ [0.5ex] 
 \hline\hline
 0.01 & 0.0026 & 0.0028  \\ 
 \hline
 0.05 & 0.0130 & 0.0142  \\
 \hline
 0.10 & 0.0261 & 0.0284  \\
 \hline
 0.15 & 0.0391 & 0.0425  \\[1ex] 
 \hline
\end{tabular}
\end{subtable}
\hfill
\begin{subtable}[b]{0.46\textwidth}
 \centering
 \begin{tabular}{||c c c||} 
 \hline
 $\alpha$ & $\eta_{\mathrm{int}}$ & $\eta_{\mathrm{non-int}}$  \\ [0.5ex] 
 \hline\hline
 0.01 & 0.0020 & 0.0019  \\ 
 \hline
 0.05 & 0.0101 & 0.0096  \\
 \hline
 0.10 & 0.0203 & 0.0192  \\
 \hline
 0.15 & 0.0305 & 0.0289  \\[1ex] 
 \hline
\end{tabular}
\end{subtable}
\end{center}
\caption{Table for comparison of the slopes for the initial growth of the diagonal coefficients $|a_n|$ with (left) fixed $\alpha=0$ and various $\gamma$, (right) fixed $\gamma=0$ and various $\alpha$.}
\label{table1}
\end{table}

\begin{figure}[t]
   \centering
\begin{subfigure}[b]{0.46\textwidth}
\centering
\includegraphics[width=\textwidth]{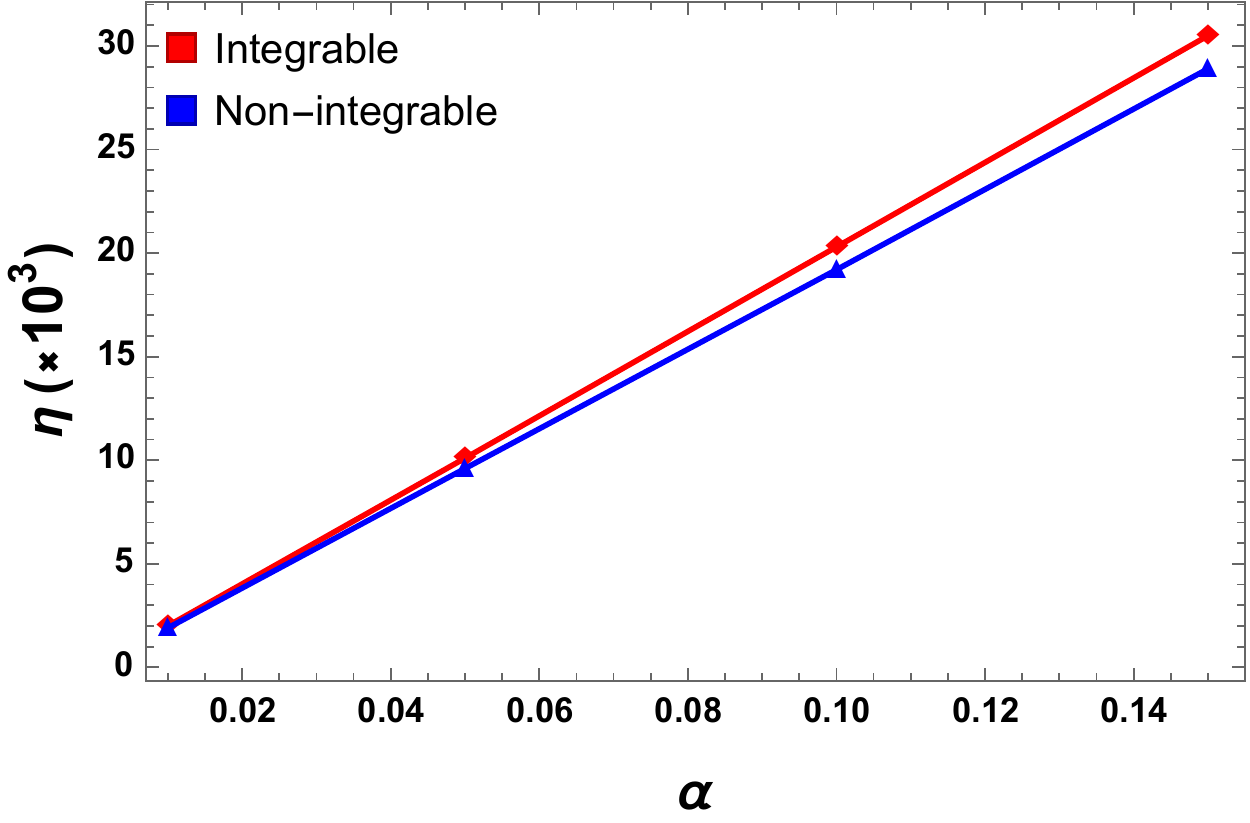}
\caption{ $\eta$ vs $\alpha$.}
\end{subfigure}
\hfill
\begin{subfigure}[b]{0.46\textwidth}
\centering
\includegraphics[width=\textwidth]{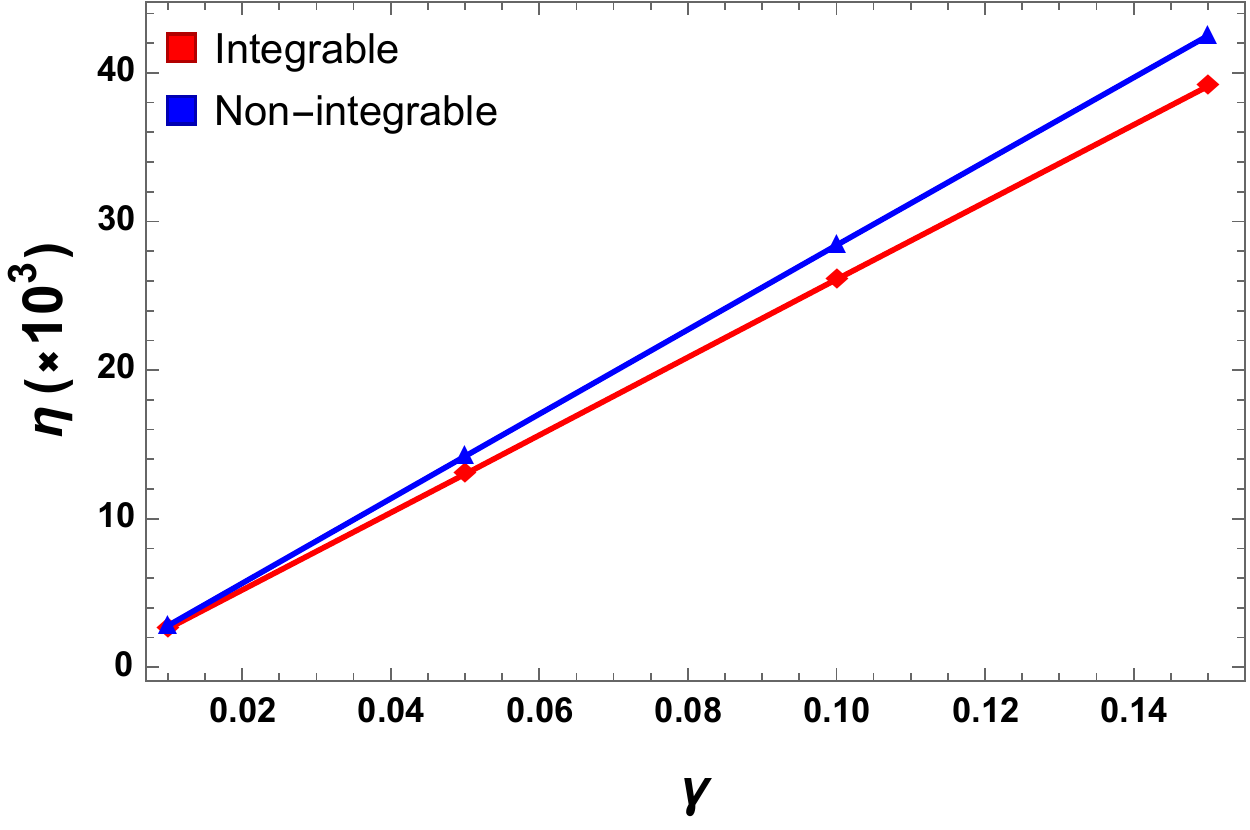}
\caption{$\eta$ vs $\gamma$.}
\end{subfigure}
\caption{Behavior of $\eta$ with $\alpha$ and $\gamma$, according to Eq.\eqref{eta}. In both cases, the plots show linear growth. We use values from Table\,\ref{table1} (a) and (b).} \label{fig:etaslopes}
\end{figure}

\begin{itemize}
    \item Initially, the integrable and non-integrable Lanczos sequences are indistinguishable from the closed system, which is manifested in the first few Lanczos coefficients. However, the distinguishability appears in the later and larger $n$. We find that, in this regime,
    similar to the integrable one (Fig.\,\ref{fig:bnopenTFIM} (a)), there exists a substantial amount of fluctuations\footnote{Apart from the usual fluctuations that were present in a closed system, there exists a substantial number of outliers (very large fluctuations) in the plot, which we get rid of by implementing a maximum cut-off by hand.} in the non-integrable case (Fig.\,\ref{fig:bnopenTFIM} (b)), unlike the closed system analysis. It is important to note that in the closed system studies, the lesser amount of fluctuations in the non-integrable case compared to the integrable case made the chaotic complexity saturate at a higher value than the integrable ones. Therefore, the open system Lanczos sequence suggests that at later times, the distinguishability between the integrable and chaotic regimes is lost.

    \item The purely imaginary diagonal elements of the Lindbladian initially grow with $n$ and then saturate (Fig.\,\ref{fig:bnopenTFIM} (c)). The saturation appears due to the finite size of the system. The behavior of these elements is almost similar for integrable and non-integrable cases. 
    
    \item The initial growth is linear and seems to be universally true for both integrable and non-integrable cases. In general, the behavior of the initial diagonals before reaching saturation is the following
    \begin{equation}
        |a_n|\sim \eta \,(\alpha, \gamma) \,n + k\,,
    \end{equation}
    with $\eta\,(\alpha,\gamma)$ is some constant and depends on the environmental couplings. Here $k$ is some offset which can be set to zero. In the Table\,\ref{table1} a) and b), we compare the slopes between integrable and non-integrable cases for $i)$ fixed $\gamma$ with varying $\alpha$ and $ii)$ fixed $\alpha$ with varying $\gamma$. Although there is no straightforward proof of this statement, a probable reason is due to the operator size concentration, i.e., $n$-th Krylov basis is a linear combination of size $n$ (i.e., Pauli basis supported on $n$ sites). In large $q$ SYK, this property has been proven to hold for any generic dissipation \cite{Bhattacharjee:2022lzy, Bhattacharjee:2023uwx}.
    
    \item From, Fig.\,\ref{fig:etaslopes} (a) and Fig.\,\ref{fig:etaslopes} (b), we observe that the behavior of $\eta \equiv \eta\,(\alpha,\gamma)$ is linear with both $\alpha$ and $\gamma$. Therefore, we find
    \begin{equation}
        \eta= c_1 \alpha +c_2 \gamma\,, \label{eta}
    \end{equation}
    where $c_1$ and $c_2$ are some constants (depending on the system size) and can be obtained by the linear fit of the data. For our purposes, the slopes are not important, although we see that $c_1$ higher for the integrable regime, while $c_2$ is higher in the chaotic regime. This linearity is consistent with the results of the dissipative SYK model \cite{Bhattacharjee:2022lzy}.
\end{itemize}

The above behavior results in a decay of the probability. We plot the normalized K-complexity (Eq.\eqref{nKr}) and observe the following behavior.
\begin{figure}[t]
   \centering
\begin{subfigure}[b]{0.3\textwidth}
\centering
\includegraphics[width=\textwidth]{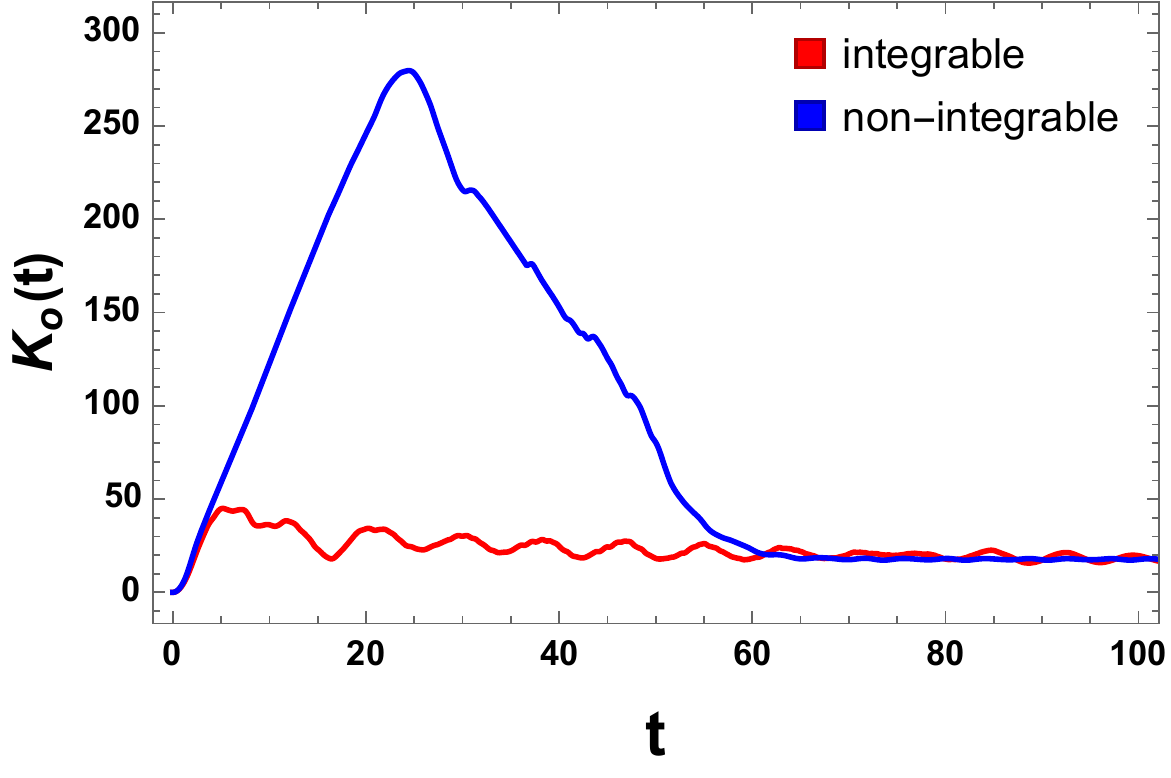}
\caption{}
\end{subfigure}
\hfill
\begin{subfigure}[b]{0.3\textwidth}
\centering
\includegraphics[width=\textwidth]{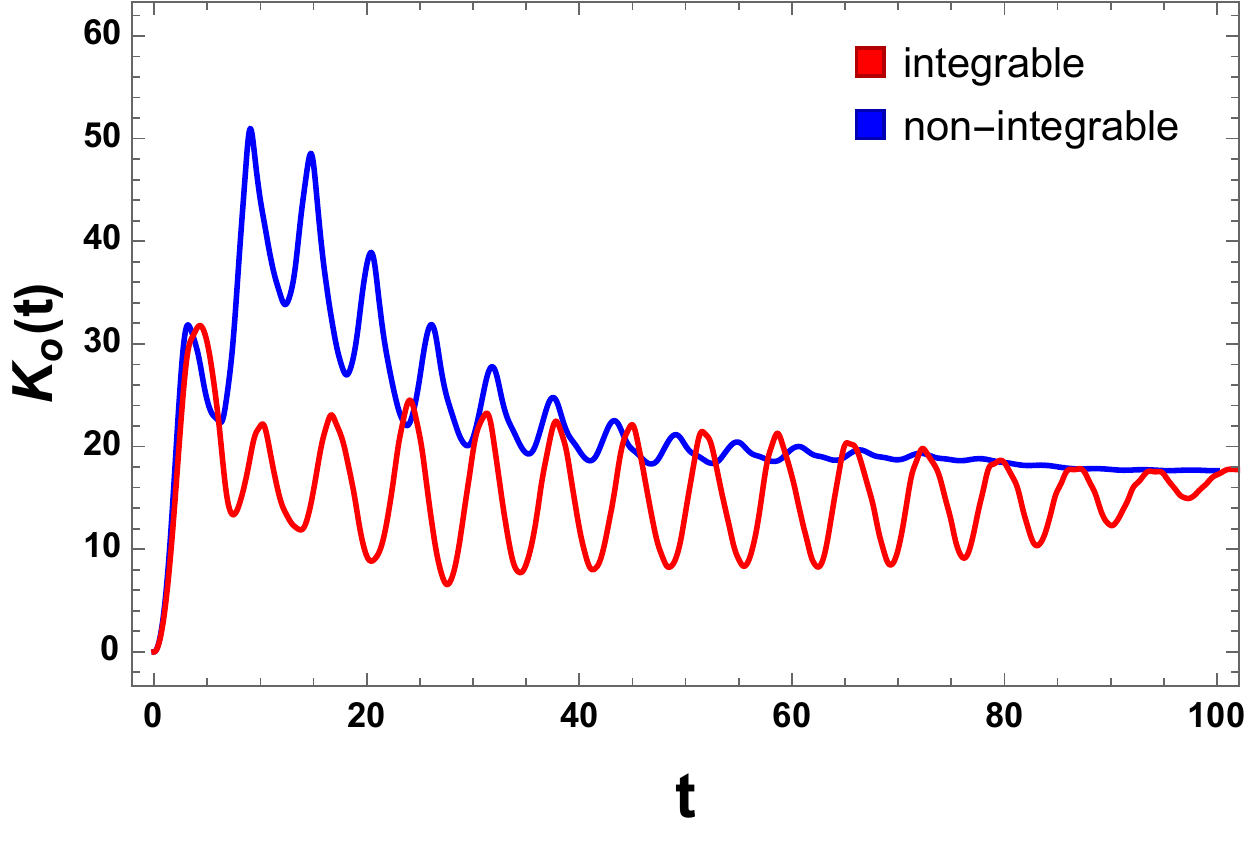}
\caption{}
\end{subfigure}
\hfill
\begin{subfigure}[b]{0.3\textwidth}
\centering
\includegraphics[width=\textwidth]{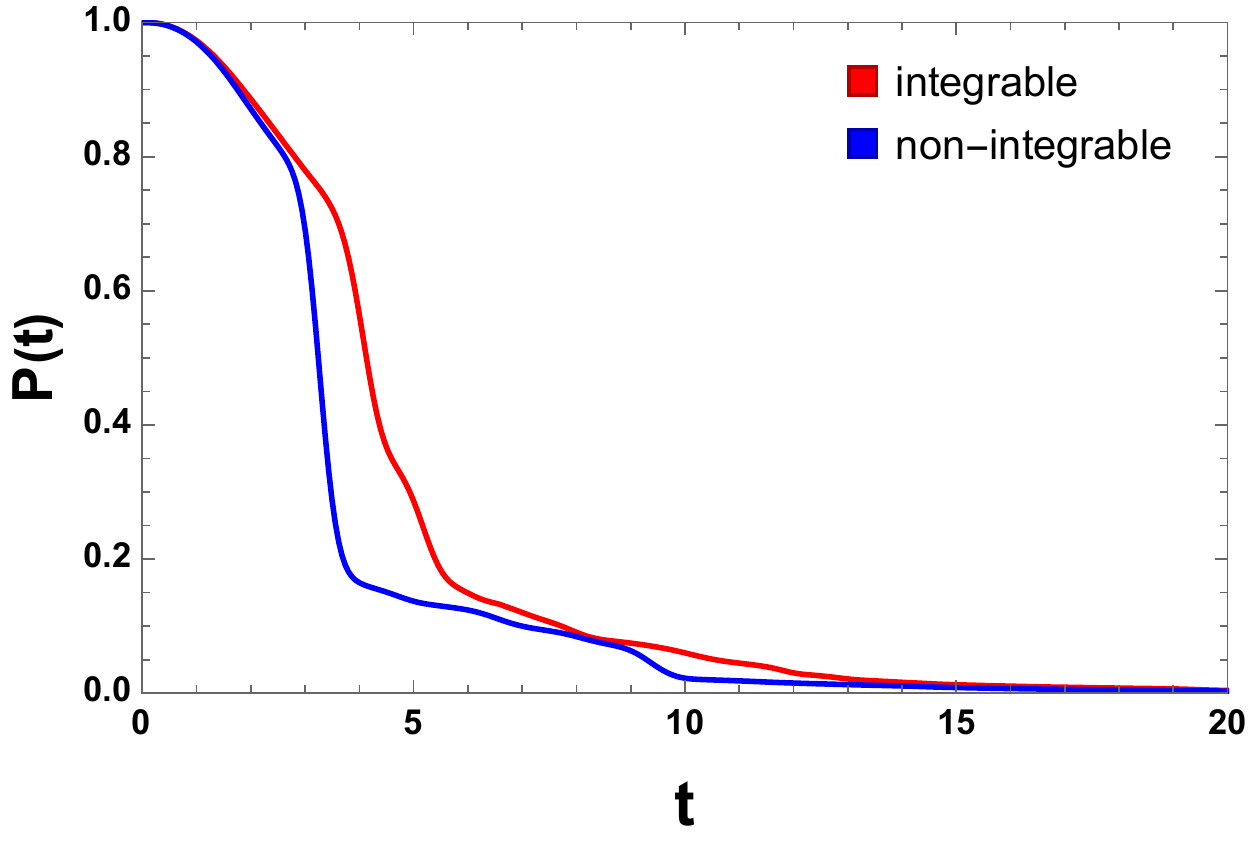}
\caption{}
\end{subfigure}
\caption{Comparison of K-complexities between integrable ($g=1$, $h=0$, red) and non-integrable ($g=-1.05$, $h=0.5$) limit for fixed $\gamma=0.01$ with (a) $\alpha=0.01$ and (b) $\alpha=0.1$. (c) shows the decay of probabilities for $\gamma=0.01$ and $\alpha=0.01$.} \label{fig:LanopenTFIM}
\end{figure}
\begin{figure}[t]
   \centering
\begin{subfigure}[b]{0.3\textwidth}
\centering
\includegraphics[width=\textwidth]{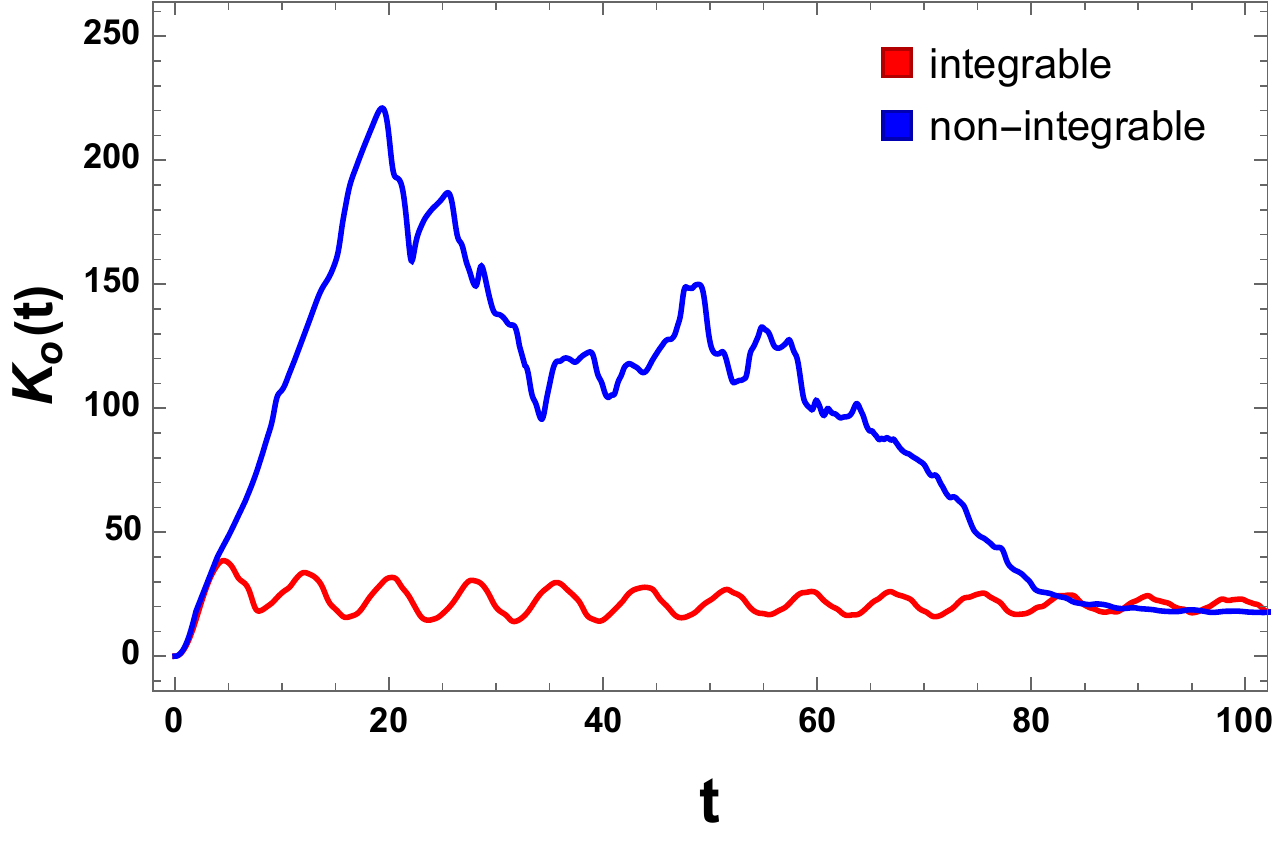}
\caption{}
\end{subfigure}
\hfill
\begin{subfigure}[b]{0.3\textwidth}
\centering
\includegraphics[width=\textwidth]{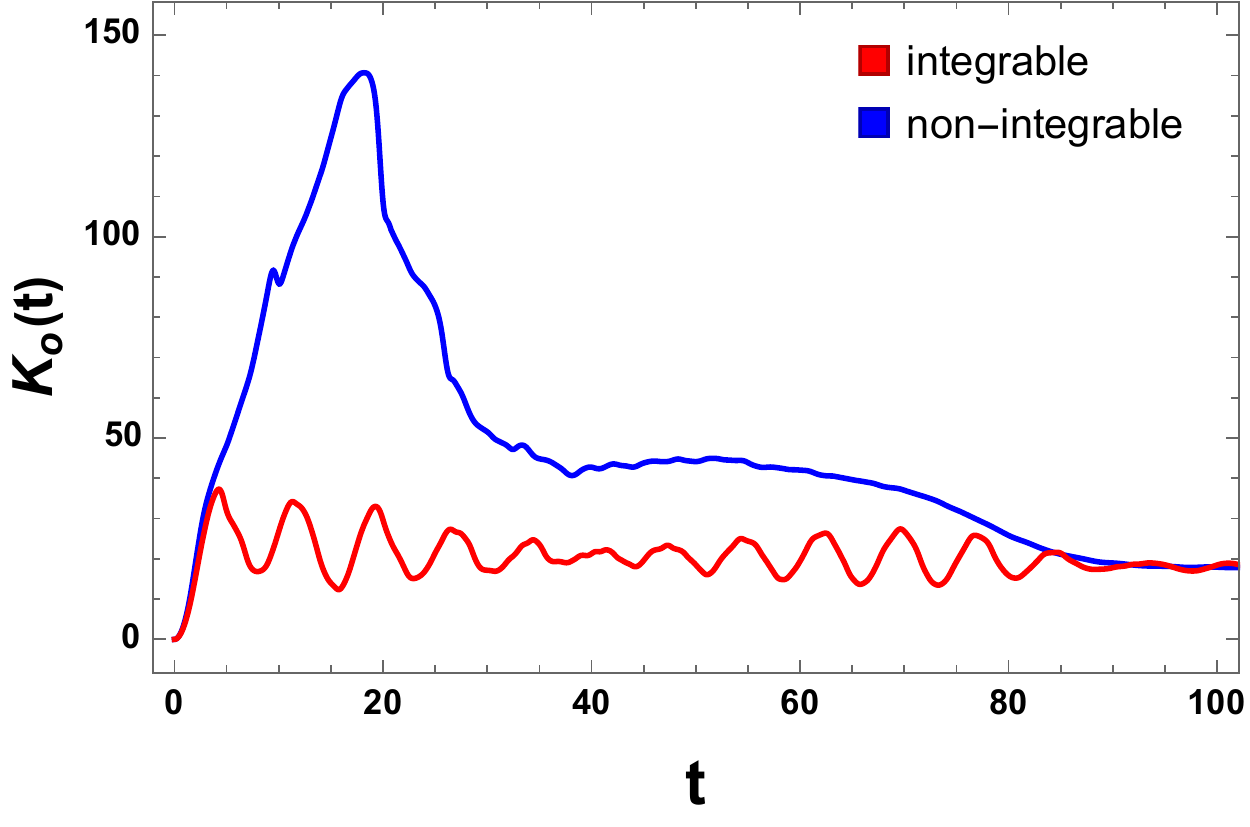}
\caption{}
\end{subfigure}
\hfill
\begin{subfigure}[b]{0.3\textwidth}
\centering
\includegraphics[width=\textwidth]{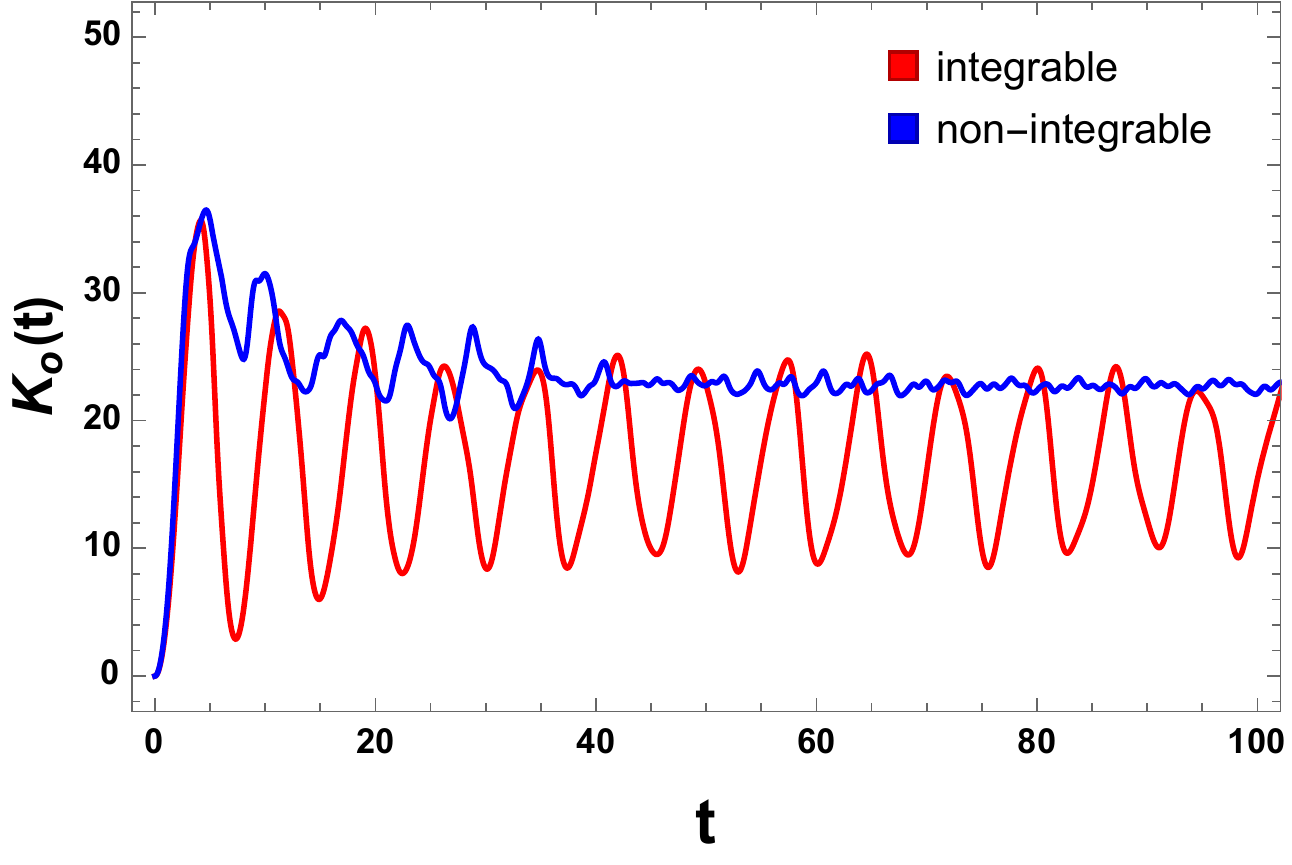}
\caption{}
\end{subfigure}
\caption{Comparison of K-complexities between the integrable ($g=1$, $h=0$, red) and the non-integrable ($g=-1.05$, $h=0.5$) for fixed $\alpha=0.05$ with (a) $\gamma=0$ and (b) $\gamma=0.01$ and (c) $\gamma=0.05$.} \label{fig:LanopenTFIM2}
\end{figure}
\begin{itemize}
    \item The K-complexity for both integrable and non-integrable cases grows initially before showing decay and saturation to a constant value (Fig.\,\ref{fig:LanopenTFIM} and Fig.\,\ref{fig:LanopenTFIM2}). The initial growth of complexity is more pronounced for the non-integrable case than the integrable case (Fig.\,\ref{fig:LanopenTFIM} (a) and Fig.\,\ref{fig:LanopenTFIM2} (a)). As we increase the environmental coupling parameters $\alpha$ and/or $\gamma$, the initial peak (before the decay) value decreases for the chaotic case (Fig.\,\ref{fig:LanopenTFIM} (b) and Fig.\,\ref{fig:LanopenTFIM2} (b) and (c)). The integrable peak value also decreases, but the decay for the integrable case is far less pronounced than the non-integrable ones.
    
    \item The late-time saturation value of the complexity seems to be universal in open systems. This comes from the fact that the late-time complexity is governed by the Lindblad jump operators while the early-time growth is controlled by the Hamiltonian.\footnote{The late-time saturation might be described by random matrix theory (RMT) universality and Ginibre ensemble \cite{Ginibre:1965zz, schomerus2017random, PhysRevLett.123.090603}. We thank Shinsei Ryu for the discussions on this point.} The saturation appears to be the same for the integrable and non-integrable cases and does not change with increasing $\alpha$ or $\gamma$. For larger environmental couplings, the initial peak of complexity for the integrable and the non-integrable cases becomes almost the same. 
\end{itemize}

Apart from the numerical observations mentioned above, there are a couple of facts worth mentioning here. It is known that the eigenvalues of the Lindbladian ($i \mathcal{L}_o$) come in complex conjugate pairs, with their real part being always negative \cite{Bhattacharya:2022gbz}. Therefore the non-unitary evolution of the initial operator through $e^{i\mathcal{L}_o t}O(0)=e^{\Tilde{\mathcal{L}}_o t} O(0)$ always gives rise to a decay, which is known in mathematics as controlled systems. For such systems, various physical observables are known to be bounded and never reach infinity. The criterion of our system to be a controlled system is therefore dictated by the eigenvalues of the Lindbladian which can be explicitly checked through the Routh-Hurwitz stability criterion \cite{routh1877treatise, Hurwitz1895}. We have checked this with our tridiagonal form with the numerical values, and find that the criterion is satisfied.

\subsubsection{Finite-size effect and the choice of the initial operator}\label{sizesystem}

In the previous section, we studied the TFIM model with the number of sites $N=6$ and initial operator $O(0)$ having nontrivial support only in the third site, i.e., 
\begin{align}
    O(0)=\sigma_3^z \equiv \mathcal{I}\otimes \mathcal{I}\otimes \sigma^z\otimes \mathcal{I}\otimes\mathcal{I}\otimes \mathcal{I}\,.
\end{align}
For this case, the upper bound for the Krylov space dimension is $\mathcal{K}\leq(D^2-D+1)\approx 4000$, with $D=2^6$ being the Hilbert space dimension. This small size restricts us to conclude whether the observed properties are universal or simply due to the finite-size effect. To remedy this, we increase the system size and observe the generic trend of the bi-Lanczos coefficients. In particular, we discuss how the initial Lanczos coefficients behave when the system size is increased. We study the following two cases: \\
\newline
\textbf{Case 1:} System size $N=8$ and the initial operator having nontrivial support at the fourth site.\\
\newline
\textbf{Case 2:} System size $N=10$ and the initial operator having nontrivial support at the fifth site.\\

Now, we restrict our study in these higher system size cases only to the initial few Lanczos coefficients due to the limitations of numerical resources, thus we resist commenting on the finite-size effect on late-time behavior of the K-complexity. To be more specific, for $N=8$, the Krylov space dimension turns out to be $\mathcal{K}\approx 65000$, and for $N=10$, $\mathcal{K}\approx 10^6$. However, as we will see, we can get a good understanding of how the K-complexity would behave for higher system size from these initial Lanczos coefficients. 
\begin{figure}[t]
   \centering
\begin{subfigure}[b]{0.46\textwidth}
\centering
\includegraphics[width=\textwidth]{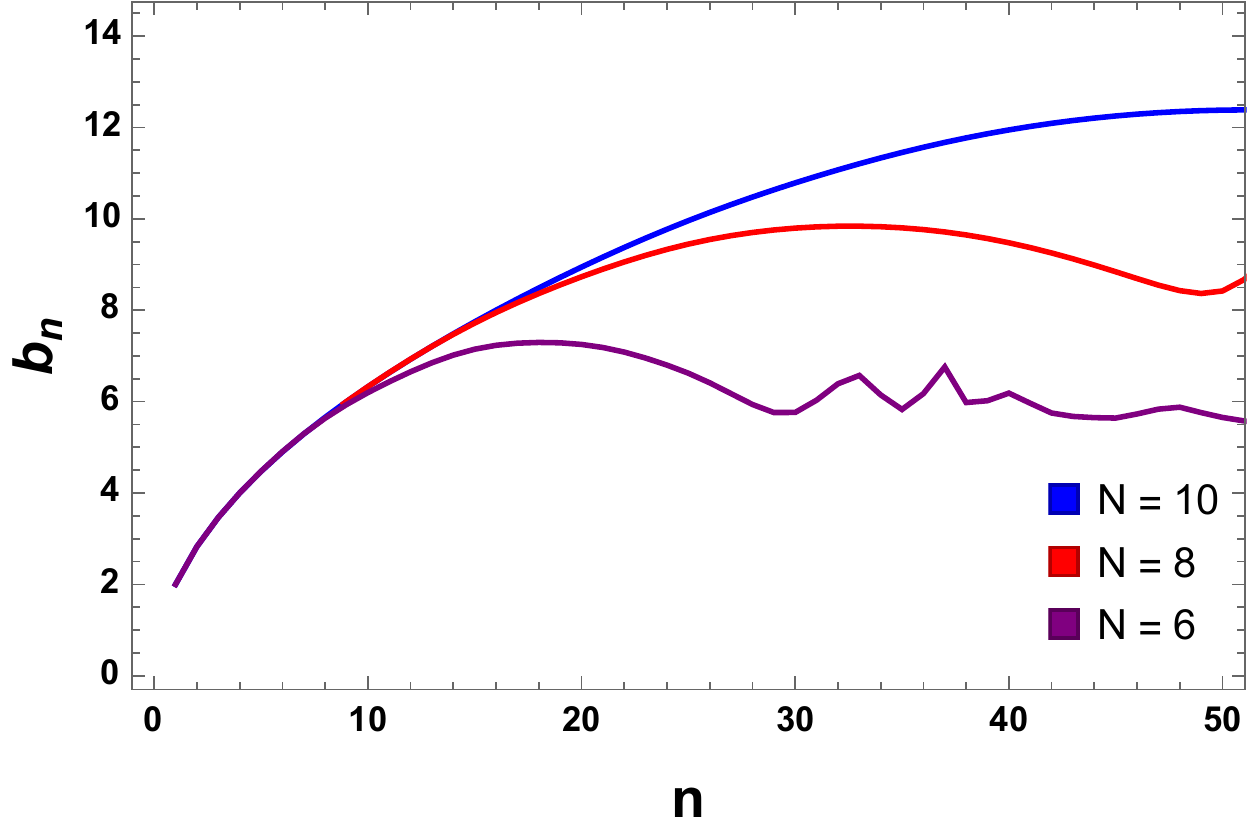}
\caption{}
\end{subfigure}
\hfill
\begin{subfigure}[b]{0.46\textwidth}
\centering
\includegraphics[width=\textwidth]{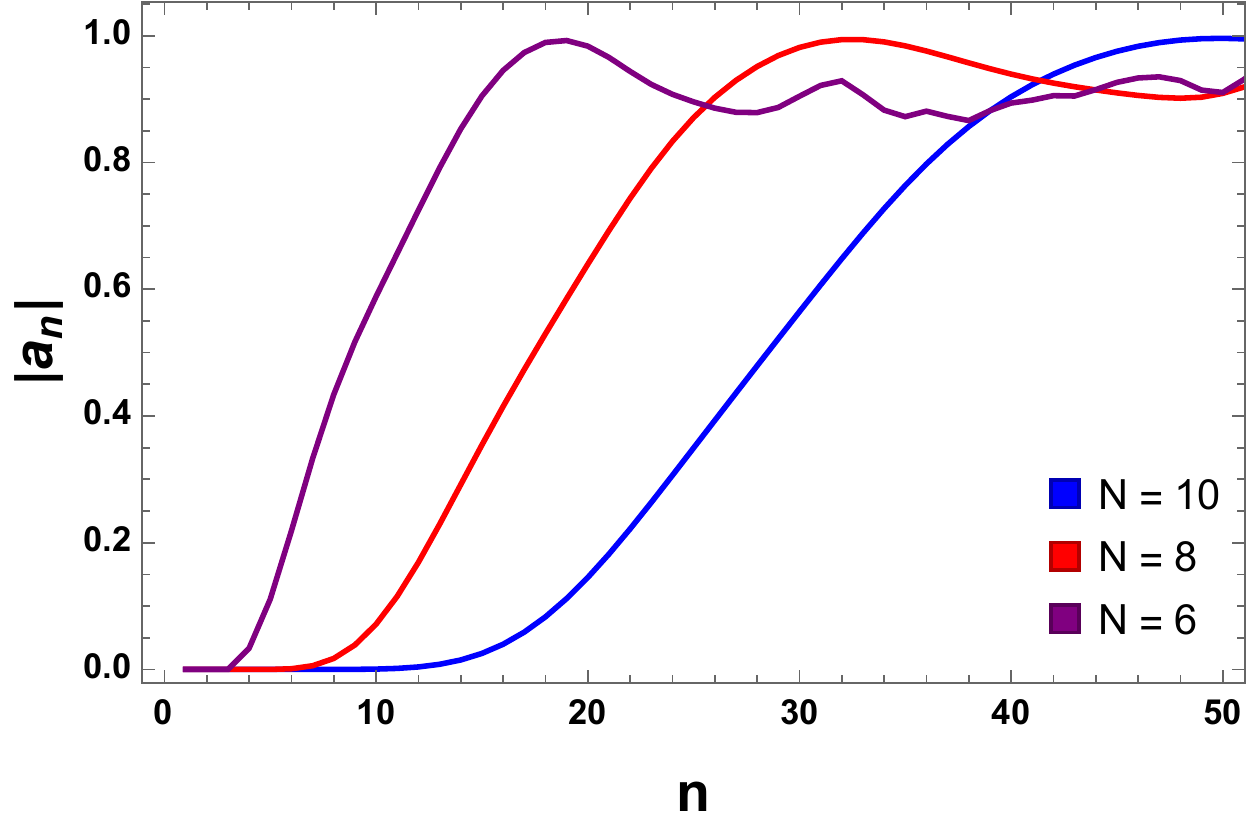}
\caption{}
\end{subfigure}
\caption{(a) Growth of initial Lanczos coefficients $b_n$ in the integrable ($g=1$, $h=0$) limit by implementing the bi-Lanczos algorithm for $N=6$ (purple), $N=8$ (red), and $N=10$ (blue) system sizes, with the probe initial operators spaced at $\sigma_3^z$, $\sigma_4^z$ and $\sigma_5^z$ positions respectively. (b) The growth of diagonal Lanczos coefficients for integrable ($g=1$, $h=0$) limit. $N=6$ (purple), $N=8$ (red), and $N=10$ (blue) system sizes are compared with probe initial operators spaced at $\sigma_3^z$, $\sigma_4^z$ and $\sigma_5^z$ positions respectively. All plots have been done for zero bulk dephasing and the boundary dephasing is set to $\alpha=0.1$.} \label{fig:diffNint}
\end{figure}

\begin{figure}[t]
   \centering
\begin{subfigure}[b]{0.46\textwidth}
\centering
\includegraphics[width=\textwidth]{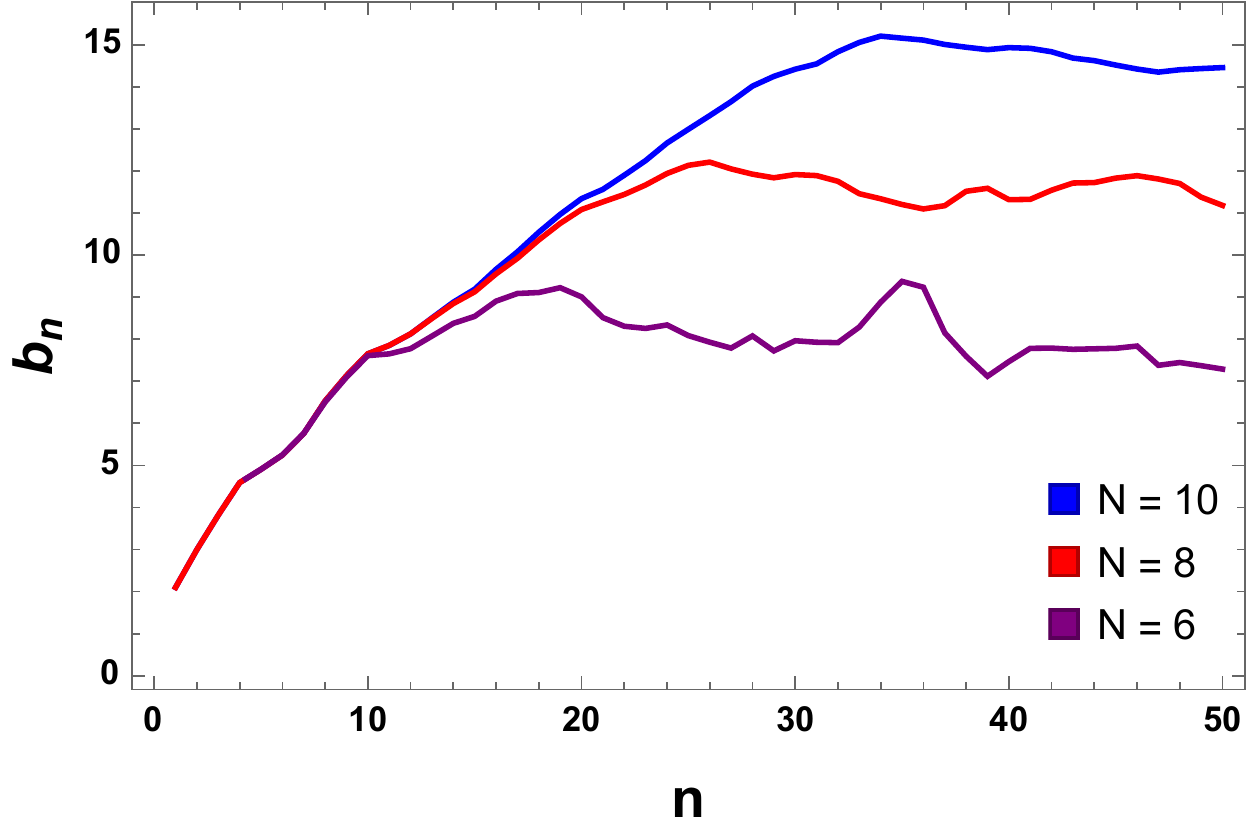}
\caption{}
\end{subfigure}
\hfill
\begin{subfigure}[b]{0.46\textwidth}
\centering
\includegraphics[width=\textwidth]{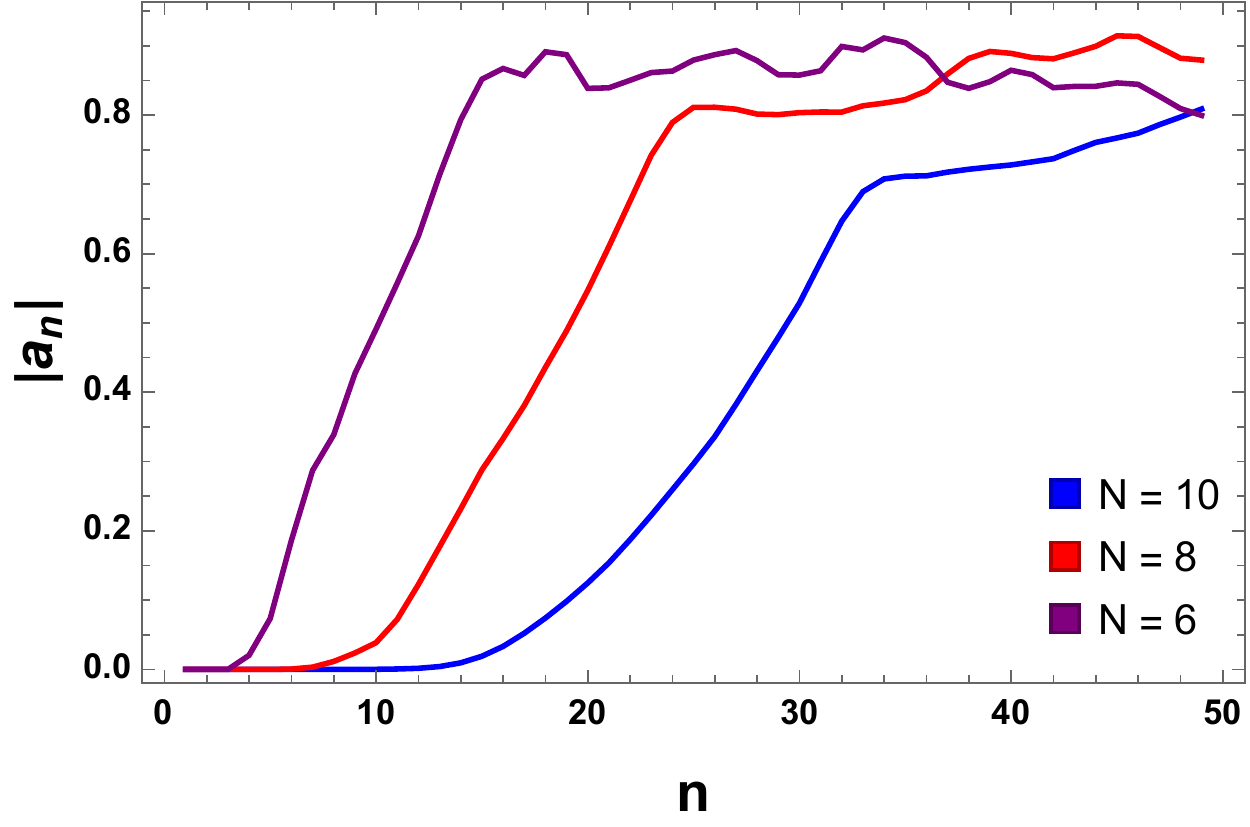}
\caption{}
\end{subfigure}
\caption{(a) Growth of initial Lanczos coefficients $b_n$ in the non-integrable ($g=-1.05$, $h=0.5$) limit by implementing the bi-Lanczos algorithm for $N=6$ (purple), $N=8$ (red), and $N=10$ (blue) system sizes, with the probe initial operators spaced at $\sigma_3^z$, $\sigma_4^z$ and $\sigma_5^z$ positions respectively. (b) The growth of diagonal Lanczos coefficients for integrable ($g=1$, $h=0$) limit. $N=6$ (purple), $N=8$ (red), and $N=10$ (blue) system sizes are compared with probe initial operators spaced at $\sigma_3^z$, $\sigma_4^z$ and $\sigma_5^z$ positions respectively. All plots have been done for zero bulk dephasing and the boundary dephasing is set to $\alpha=0.1$.} \label{fig:diffNch}
\end{figure}

In Fig.\,\ref{fig:diffNint} and Fig.\,\ref{fig:diffNch}, we compare the growth of $b_n$ and $a_n$ for different sizes in the integrable and the non-integrable limit. We observe that as we increase the system size, the growth of the $b_n$ persists for larger $n$ before the finite-size effect kicks in. Interestingly, the slopes or the general behavior of $b_n$ for the integrable and the non-integrable regimes seem to be universal before the finite-size effect hits, i.e., $b_n$ only captures the integrability of the system and increasing the system size increases its saturation value, thus corroborating the previous studies in closed SYK \cite{Jian:2020qpp} and open SYK \cite{Bhattacharjee:2022lzy}.

Similarly, we observe that as the system size is increased, the slope of the magnitude of the diagonal coefficients $a_n$ remains unchanged while the growth starts in higher values of $n$ (see Fig.\,\ref{fig:diffNint} (b) and Fig.\,\ref{fig:diffNch} (b)). This can be understood from the fact that as we increase the system size with the initial operator supported in the middle of the chain, it takes more iterations (or analogously more time) to grow and reach the boundary, where dephasing operators act. As a result, the diagonal coefficients, which are purely a result of the dissipative effect and the non-Hermiticity of the Lindbladian, start showing nonzero values later for larger system sizes. Surprisingly, the generic behavior of $a_n$ seems to be linear in both integrable and non-integrable systems and saturates due to the finite-size effect, corroborating our previous studies with Arnoldi iteration \cite{Bhattacharya:2022gbz, Bhattacharjee:2022lzy}.\footnote{The above results allow us to comment on the K-complexity at the early-time regime but not late-time. Since the initial growth is mostly controlled by $b_n$, it would persist longer for larger-sized systems.}

Further, in Fig.\,\ref{fig:diffNint} and Fig.\,\ref{fig:diffNch}, we note that there is a slight delay in the number of steps $n_1$ where $a_n$ start growing compared to the number of steps $n_2$ where $b_n$ start to saturate. The operator \emph{hitting on the wall} can be defined when the $n$-th Krylov basis has \emph{at least} support on \emph{one} operator of size $n$. This happens on step $n_1$ slightly earlier than step $n_2$ when operator \emph{completely} hits the wall. This is the step when the operator size concentration \cite{Bhattacharjee:2022lzy} holds and $b_n$ saturates. These two situations arise because of the \emph{presence of boundary dissipation only} and \emph{absence of bulk dissipation}. The bulk dissipation makes the operator decay along with its growth; hence, the $a_n$ starts to grow even from the first step.

We briefly explain the operator size concentration in this context. In the general operator Hilbert space for $N$-site Hamiltonian, we can always write the $n$-th Krylov basis operator as
\begin{equation}
    \mathcal{O}_n = \sum_{i=\{1\}}^{\{N\}} c_{\{i\}} \sigma_{\{i\}}\,,
\end{equation}
where $c_{\{i\}}$ are a class of real coefficients and $\sigma_{\{i\}}$ denotes the combination of Pauli spin operators which has support on $i$ sites. For example, $\sigma_{\{3\}}$ would mean a set of all operators having nontrivial support on \emph{any} three of the $N$ sites. In this language, the scrambling time can be characterized by the time when the $n$-th Krylov operator $\mathcal{O}_n$ is majorly dominated by the set of operators ($\sigma_{\{N\}}$) having support on the full size ($N$) of the system. Mathematically this means that at $t=t_{\mathrm{scr}}$, we have $c_{\{N\}} \gg c_{\{i\}}$, for all $i<N$. In conjunction with the previous paragraph, this happens after $n_2$ number of Lanczos iterations. On the other hand, $n_1$ corresponds to the Lanczos step when $N$ site-supported coefficients $c_{\{N\}}$ become nonzero for the first time, but the contribution need not dominate over the contributions from lesser supported operators.

\begin{figure}[t]
   \centering
\begin{subfigure}[b]{0.46\textwidth}
\centering
\includegraphics[width=\textwidth]{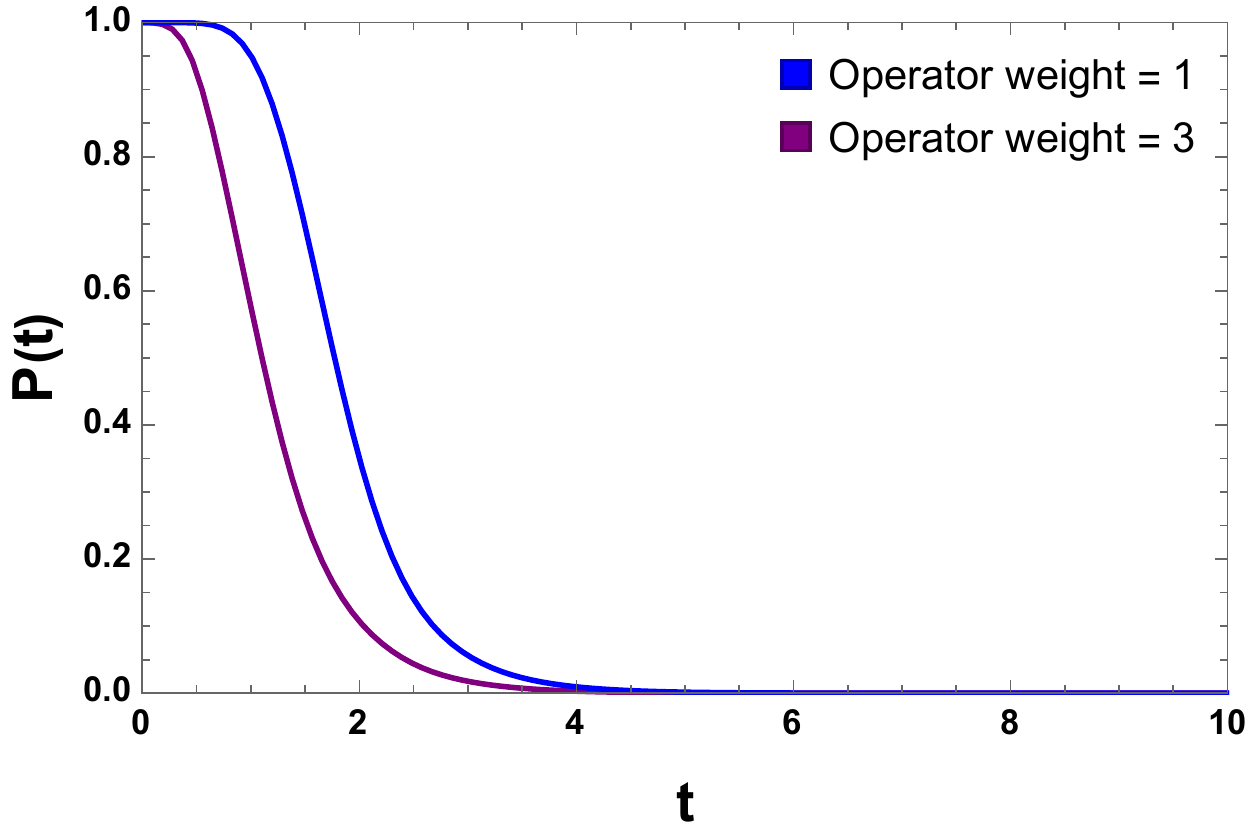}
\caption{}
\end{subfigure}
\hfill
\begin{subfigure}[b]{0.46\textwidth}
\centering
\includegraphics[width=\textwidth]{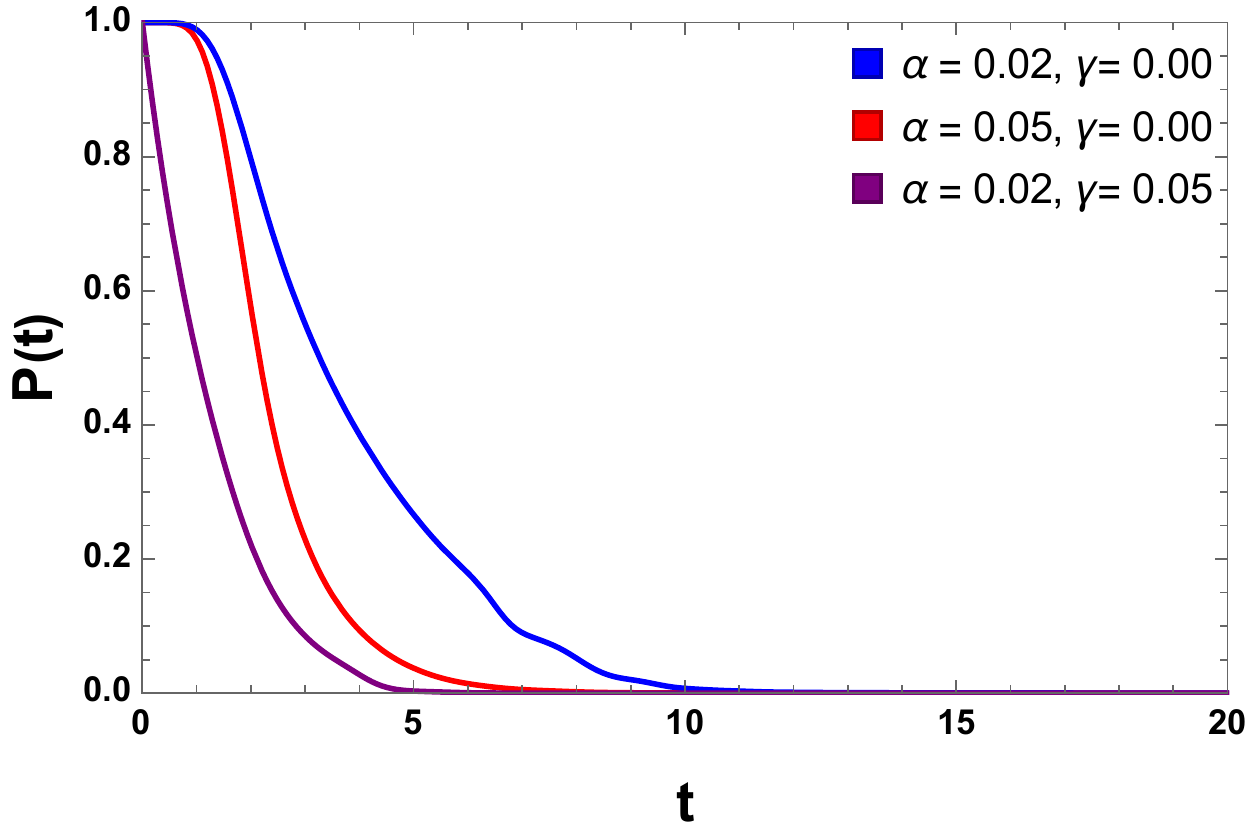}
\caption{}
\end{subfigure}
\caption{(a) Comparison of the decay of total probabilities for different weights (operator weight one $\sigma_3^z$ (blue) and three $\sigma_2^z\otimes \sigma_3^z\otimes \sigma_4^z$ (purple) respectively) of the initial operator in the integrable ($g=1$, $h=0$) limit. Here the Lindblad couplings are $\gamma = 0$ and $\alpha=0.1$. (b) Comparison of the decay of total probabilities for different boundary dephasing couplings $\alpha=0.02,\, \gamma=0$ (blue), and $\alpha=0.05,\, \gamma=0$ (red) and $\alpha=0.02, \, \gamma=0.05$ (purple) in the non-integrable ($g=-1.05$, $h=0.5$) limit. We choose the initial operator $\sigma_3^z$. In both cases, the system size is $N=6$.} \label{fig:diffspread}
\end{figure}

We also discuss the results of the increasing support of the initial operator for a given system size $N=6$. It can already be intuitively understood from our results - once the support of the initial operator is increased from one site to many, the operator becomes more non-local and would come across the boundary dissipation with less number of iterations. As a result, $a_n$ would start showing nonzero values in a smaller number of iterations.

We compare the total probabilities for different weights of the initial operators. From Fig.\,\ref{fig:diffspread} (a), we can see that as the weight of the initial operator is increased, the total probability starts decaying earlier. This agrees with the above-mentioned intuition that with increasing initial operator weight, the operator evolution is exposed to the environment quicker. However, the rates of decay seem comparable to each other. As a result of this, the complexities would also start decaying earlier with increasing operator weight. Therefore the peak of the complexity seems to get lower for increased operator size since the decay regime will kick in earlier.

In addition, in Fig.\,\ref{fig:diffspread} (b), we show the time evolution of probability in the non-integrable regimes with different coupling strengths for the Lindbladian jump operators. While the blue and red plots suggest that with increasing boundary dephasing, the rate of probability decay becomes more, the purple plot shows that bulk dephasing in addition to the boundary dephasing exposes the system to an even stronger non-Hermiticity (the interaction with the environment) that results in an even quicker decay of total probability. It is worth noting that the decay rate only depends on the environmental coupling but not on the operator weight.

Finally, we make a comment on the thermodynamic limit. In this limit ($N\rightarrow \infty$), the behavior shown in Fig.\,\ref{fig:diffNint} and Fig.\,\ref{fig:diffNch} suggests that the operator will take a large number of iterations to reach the boundary where the dissipation acts. Hence, the $a_n$ coefficients will always vanish. This in turn means that open quantum systems in thermodynamic limits with only boundary dissipation behave very similar to closed systems characterized by asymptotically growing $b_n$ coefficients and no substantial effect of $a_n$ coefficients. Just like the closed systems in the thermodynamic limit, the scrambling time then tends to infinity and the initial growth of K-complexity persists. This situation will dramatically change when we introduce all-site bulk dephasing. The dissipation will be present even in the thermodynamic limit, and both $a_n$ and $b_n$ will grow depending on the integrability of the system. However, for chaotic systems, we expect both coefficients to show linear growth. This will make the K-complexity saturate at a time that decreases logarithmically with increasing dissipation strength \cite{Bhattacharjee:2022lzy, Bhattacharjee:2023uwx}.


\section{Conclusion and outlook}\label{conc}

In this paper, we extended our previous studies of operator growth and K-complexity for open quantum systems. We have considered two models, the open TFIM and the open XXZ spin chains. The main text contains the results for the open TFIM whereas similar results for the open XXZ chain are given in Appendix \ref{appb}. For both models, we implement the bi-Lanczos algorithm to construct the tridiagonal matrix representation of the Lindbladian.
 The Krylov space dimension in both cases is of similar order $(\sim 4000)$ for spin size $N = 6$. Our main conclusions of the paper are listed below.\\
\newline
\textbf{Early time and small $n$ behavior:}
The behavior of the coefficients present in the tridiagonal representations for a small number of iterations (small $n$) controls the early time behavior of the K-complexity.

\begin{itemize}
    \item The bi-Lanczos algorithm reduces to the closed system Lanczos algorithm when the environmental coupling parameters $\alpha$ and $\gamma$ are taken to be zero. This can be understood in the level of the algorithm itself, as explained in Appendix \ref{appa}.
    
    \item For zero environmental couplings, we get back the closed system K-complexity plots which agree with previous studies (Fig.\,\ref{fig:Lanclosed}). The initial growth rate of complexity is more pronounced for the non-integrable case compared to the integrable ones. This can be traced back to the initial growth rate of the Lanczos coefficients.
    
    \item For nonzero environmental couplings, the bi-Lanczos algorithm recasts the Lindbladian into a purely tridiagonal matrix. This is to be contrasted with the Arnoldi iteration, where the Lindbladian is obtained in the upper Hessenberg form \cite{Bhattacharya:2022gbz, Bhattacharjee:2022lzy}. 
    
    \item The diagonal coefficients of the Lindbladian are purely imaginary and nonzero, unlike the closed system case, where the diagonal elements vanish. These purely imaginary diagonals cause decay in the amplitudes $\phi_n(t)$, which in turn results in the decay of the probability distribution and K-complexity for both the integrable and non-integrable cases (Fig.\,\ref{fig:LanopenTFIM} (c)).
    
    \item For small $n$, the magnitude of the diagonal coefficients shows a linear growth, which is similar for both integrable and non-integrable regimes. Therefore, unlike the off diagonals $b_n$, one can not distinguish between integrable and non-integrable regimes by observing the diagonal coefficients. The slope appears to be increasing with the increasing environmental coupling parameters.
    
    \item In the off-diagonal coefficients $b_n$ for small $n$ (up to $50$), we see exactly similar behavior as the closed system. Hence the distinguishability between integrable and non-integrable cases remains intact initially, resulting in a distinguishability in the early growth of the K-complexity.
    
    \item Once the environmental coupling parameters are increased, the growth rate of the magnitudes of the diagonal coefficients increases, resulting in a loss of early-time distinguishability between the integrable and non-integrable cases. This essentially means that with increasing environmental couplings, the information on integrability washes out quickly. 

    \item We also compare the small $n$ behavior of the Lanczos coefficients for different system sizes, with different weights of initial operators. The finite-size effect has a strong impact on the saturation of $b_n$, especially the saturation value increases with system sizes. A similar conclusion can be drawn for $a_n$ as well but here the saturation value remains comparable for different system sizes while the saturation happens at higher $n$ values for larger system sizes. This behavior appears to be ``universal'' in the sense that the off-diagonal elements of the Lindbladian are sensitive to the integrability of the system while the diagonal elements ($a_n$) capture the dissipation. 
    
    \item It is also worth noting that for boundary dissipation, the appearance of diagonal coefficients is delayed as the system size is increased. From this we conjectured that for boundary dissipation in the thermodynamic limit, the diagonal coefficients would take forever to appear and K-complexity will behave very similar to the closed systems.  
\end{itemize}
Therefore, the main conclusion for early times is that the integrable and the non-integrable regimes are distinguishable in the early times through the initial growth of the K-complexity only for small environmental couplings.\\
\newline
\textbf{Late time and large $n$ behavior:}
In the following, we address the late-time distinguishability between integrable and non-integrable regimes, which is controlled by the large $n$ behavior of the coefficients $a_n$ and $b_n$.

\begin{itemize}
    \item The diagonal coefficients saturate after an early-time linear growth. This saturation value, for boundary dissipation, is independent of system size and depends on the non-hermitian coupling. 
    
    \item The off-diagonal coefficients, however, exhibit a bi-Lanczos descent until the system Krylov space is fully explored. 
    
    \item The off-diagonal coefficients, even for zero environmental couplings show more fluctuations for the integrable case, as was found previously in \cite{Rabinovici:2022beu}, compared to the non-integrable case, resulting in a lower saturation compared to the non-integrable case after the initial growth. 
    
    \item For nonzero environmental couplings, both the integrable and non-integrable $b_n$ for large $n$ show even more fluctuations than the closed case. Since these later fluctuations control the saturation value of the K-complexity, one would expect that the integrable and the non-integrable cases should saturate in similar values, given the diagonals also behave similarly for the two cases in open system analysis. We indeed find the expected late-time behavior of the K-complexity, i.e.,  the saturation values are almost indistinguishable for the integrable and non-integrable cases.
 
    \item The saturation appears to be universal for a given Krylov dimension (we expect it to increase if the system size is increased, see \cite{Bhattacharjee:2022lzy} for dissipative SYK). The fact that after this saturation the complexity remains saturated for a long time, indicates that this might be a result of reaching a steady state. This in general depends both on the system size and the dissipation strength \cite{Shirai_2020}.
    
\end{itemize}
 Therefore we conclude that the notion of late-time chaos becomes completely unclear due to more fluctuations coming from Lanczos coefficients, even for chaotic evolution. Unlike the early-time behavior, the late-time saturation value of the K-complexity is always the same. This is not only independent of the nature of the Hamiltonian (integrable or chaotic) but also independent of the environmental couplings as long as they are non-vanishing. 

Hence, it is clear that the information of integrability is there at early times for only small enough dissipation. However, the distinguishability is always lost at late times for open system dynamics, even for small dissipation. This indicates that however small the dissipation might be, it washes out the information of integrability as we go further in time and makes the complexity saturation value independent of the dissipation or integrability. Therefore, at late times, all the cases studied in our paper for a given model, have almost the same complexity. We, therefore, expect this saturation value to be an increasing function of the system size. It would be however interesting to understand this behavior analytically. We hope to explore this direction in future works.

We conclude with a few interesting future directions for this work. Firstly, given the recent ventures of computing Krylov complexity in quantum field theories \cite{Dymarsky:2021bjq, Avdoshkin:2022xuw, Camargo:2022rnt}, it would be interesting to study the thermal autocorrelation functions and Krylov complexity in open QFTs \cite{ Loganayagam:2022zmq}. The non-unitary evolution in such a case can be modeled by the Feynman-Vernon influence functional along with the Schwinger-Keldysh formalism \cite{Baidya:2017eho}. Further, the bi-Lanczos algorithm is applicable not only to operator growth but also to any non-Hermitian evolution. For example, it should apply equally well in the studies of the spread complexity in the Schr\"odinger picture given the Hamiltonian under which the evolution happens, is non-Hermitian (see Appendix \ref{biL_spread}). A few of such scenarios, where a non-Hermitian Hamiltonian evolution occurs, are $i)$ the evolution of mixed state density matrix in terms of an effective non-Hermitian Hamiltonian \cite{Matsoukas-Roubeas:2022odk, Alishahiha:2022anw} derived from the Lindblad master equation and $ii)$ the evolution of quantum states under projective measurements after regular intervals (also known as the first passage problem) \cite{PhysRevA.91.062115}. In the former case, it might be interesting to understand the Liouvillian gap and the corresponding relaxation time scale from the operator growth perspective for integrable systems \cite{Shibata:2018bir, Shibata:2019cvd, Shibata:2020cii}.\footnote{We thank Hosho Katsura for discussions regarding this point.}  We hope to address some of the questions in future studies.

\section*{Acknowledgements}
We wish to thank Xiangyu Cao, Rathindra Nath Das, Bidyut Dey, Hosho Katsura, Amin Nizami, Tanay Pathak, Shinsei Ryu, Aninda Sinha, Julian Sonner, and Hironobu Yoshida for useful discussions in related works. We also thank the anonymous referee for valuable suggestions, which leads to the subsection \ref{sizesystem}. A.B. thanks the organizers of the workshop on ``QI in QFT and AdS/CFT III" and the hospitality of the Theory Division, Saha Institute of Nuclear Physics, where part of this work was presented. P.N. would like to thank Princeton Center for Theoretical Science (PCTS), Princeton University for hosting him through the overseas visiting program for young researchers (KAKENHI No.\,21H05182) during the final stages of the work. The work of A.B. is supported by the Polish National Science Centre (NCN) grant 2021/42/E/ST2/00234. The work of P.N. is supported by the JSPS Grant-in-Aid for Transformative Research Areas (A) ``Extreme Universe'' No.\,21H05190.

\appendix
\section{Appendix: Reduction of bi-Lanczos algorithm into Lanczos algorithm} \label{appa}

Consider the application of the Lanczos algorithm on the Hermitian matrix. In this case, we expect it to reduce to the Lanczos method. 

Firstly, the original Lanczos algorithm can be written in the following way.

\begin{enumerate}
   \item Let $|q_0 \rrangle \in \mathbb{C}^n$ be an arbitrary vector with $\lVert q_0\rVert =1$.
    \item Abbreviated initial iteration steps are the following:
    \begin{enumerate}
        \item Let $|r'_0 \rrangle  = A |q_0\rrangle $.
        \item Compute the inner product, $\alpha_0=\llangle r'_0|q_0 \rrangle $.
        \item Define $|r_0 \rrangle  =|r'_0 \rrangle  -\alpha_0 |q_0 \rrangle $.
    \end{enumerate}
    \item For $j=1, 2,\cdots,$ perform the following steps:
    \begin{enumerate}
        \item Compute the norm, $\beta_j=\lVert r_{j-1}\rVert$.
        \item If $\beta_j \neq 0$, define $|q_j \rrangle  =|r_{j-1}\rrangle /\beta_j$.
        \item Let $|r'_j \rrangle  =A |q_j \rrangle$. 
        \item Compute the inner product, $\alpha_j = \llangle q_j| r'_j \rrangle $. 
        \item Define $|r_j \rrangle  = |r'_j \rrangle  - \alpha_j |q_j \rrangle  - \beta_j |q_{j-1} \rrangle$.
        \item If $\beta_j = 0$, stop, otherwise go back to step $3$.
    \end{enumerate}
    \item Let $Q$ be the matrix with columns $q_1,\ldots ,q_m$. Then $T=Q^{*}AQ$.
    
\end{enumerate}
Here, note that this step-by-step algorithm is written so that one can make an easy comparison with the bi-Lanczos algorithm in the main text. By comparison, it is easy to see that in case the operator $A$ is Hermitian, the bi-Lanczos algorithm becomes redundant in the sense that it is a double copy of the Lanczos algorithm where every $\llangle q_n|$ vector can be identified as just the complex conjugate of $|p_n \rrangle$. Similarly $|r_j^{\prime}\rrangle= |s_j^{\prime}\rrangle$ and $b_j=c_j=\beta_j$. Also, the $\alpha_j=a_j$ coefficients become real in this case, due to the Hermiticity of the operator. More specifically, for operator complexity, this becomes zero as the Liouvillian superoperator involves commutation.


    



\section{Appendix: Results for open XXZ Hamiltonian}\label{appb}

The XXZ Hamiltonian is used to describe the behavior of a system of interacting spin-1/2 particles in a magnetic field. It has been used to study a wide range of physical systems, including quantum spin chains, and has been the subject of extensive research in condensed matter physics and quantum mechanics. The exact form of the Hamiltonian depends on the specific physical system being modeled, but it generally includes terms that describe the interactions between the spins of the particles as well as the interactions between the spins and the external magnetic field. We will consider nearest-neighbor interaction, and the Hamiltonian is given by
\begin{equation}
    H_{\mathrm{XXZ}} = \sum_{i=1}^{N-1} J\,(S_i^x S_{i+1}^x+S_i^y S_{i+1}^y) + J_{zz}\,S_i^z S_{i+1}^z\,,
\end{equation}
where $S_i^a = \sigma_i^a/2$ where $\sigma_i^a$ are he three Pauli matrices. This Hamiltonian is integrable for all values of the coupling parameters $J$ and $J_{zz}$. To break integrability, add the following integrability breaking term \cite{Rabinovici:2022beu}
\begin{equation}
    H_d = S_j^Z\,,
\end{equation}
with coupling parameter $\epsilon$. The integrability of the model depends on $\epsilon$, in particular, the level statistics of it match with the GOE ensemble for $\epsilon = 0.5$, which we will take our definition of ``chaotic'' Hamiltonian for our calculation purpose \cite{Rabinovici:2022beu}.\\
\newline
\textbf{Choice of sector:} The XXZ Hamiltonian commutes with the operator denoting the total spin of the system, given by
\begin{equation}
\mathcal{S} = \sum_{i=1}^N S_i^z\,.
\end{equation}
Additionally, it commutes with the parity operator ($\mathcal{P}$), meaning that it is invariant under reflection with respect to the edge of the chain. The symmetries of the XXZ Hamiltonian, specifically the conservation of total spin and parity, lead to degeneracy in the energy spectrum. By selecting a specific sector with a particular total spin and parity, we can remove these symmetries and reduce the problem's dimensionality. This can make it more numerically efficient to study operator growth in large systems. By breaking the symmetries, we can also gain a better understanding of the system's underlying physics, as the symmetries can obscure certain system features.\\
\newline
\textbf{Choice of initial operator:} Moreover, we will have to select the initial operator in such a way that it stays within the specific sector of the Hamiltonian throughout the entire evolution. We will choose the following from  the initial operator
\begin{equation}
    S_i^z + S_{N-i+1}^z  \,.
\end{equation}
\begin{figure}[t]
   \centering
\begin{subfigure}[b]{0.46\textwidth}
\centering
\includegraphics[width=\textwidth]{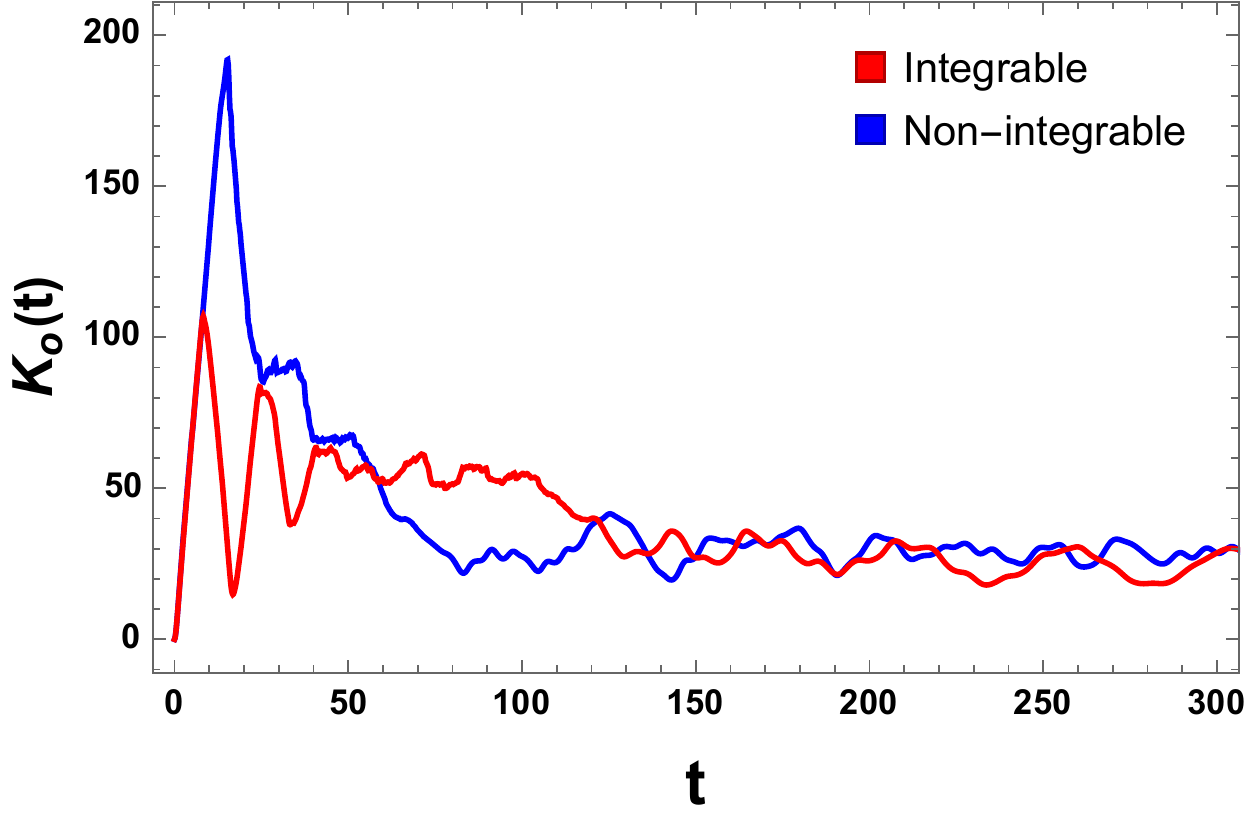}
\caption{}
\end{subfigure}
\hfill
\begin{subfigure}[b]{0.46\textwidth}
\centering
\includegraphics[width=\textwidth]{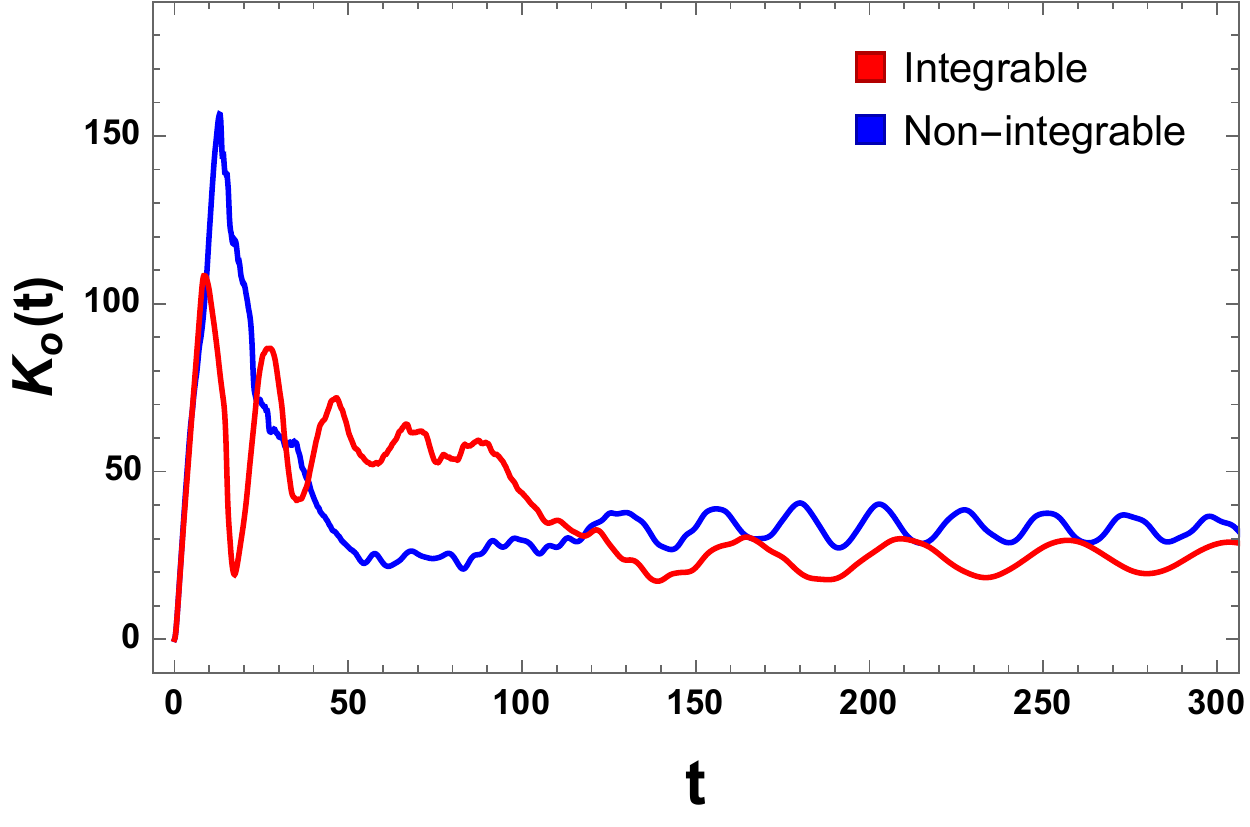}
\caption{}
\end{subfigure}
\caption{Comparison of K-complexities between integrable ($\epsilon=0$) (red) and non-integrable ($\epsilon=0.5$) (blue) for fixed $\gamma=0.01$ with (a) $\alpha=0.01$ and (b) $\alpha=0.05$ for open XXZ Hamiltonian. System size $N=16$ with total spin $\mathcal{S}=2$ and parity $\mathcal{P}=+1$, and initial operator ($S_7^z+ S_{10}^z$).} \label{fig:LanopenXXZ}
\end{figure}
\textbf{Choice of Lindblad operators:} To study the K-complexity for the open case, we will have to evolve the system with the Lindbladian instead of the Hamiltonian. We choose the Lindblad operators in such a way that they commute with the parity operator ($\mathcal{P}$) and the total spin operator ($\mathcal{S}$), only then operator will evolution will take place within the specific spin and parity sector. The bulk Lindblad operators can be chosen similarly to the TFIM case as they commute with $\mathcal{S}$ and $\mathcal{P}$. But the boundary Lindblad operators that we choose for the TFIM case do not commute with $\mathcal{P}$ and $\mathcal{S}$, and so we select the boundary operators in the following manner
\begin{equation}
     L_1 = \sqrt{\alpha}\,(\sigma_1^x\sigma_2^x + \sigma_1^y \sigma_2^y)\,, ~~~~ L_N = \sqrt{\alpha}\,(\sigma_{N-1}^x\sigma_N^x + \sigma_{N-1}^y \sigma_N^y)\,.
\end{equation}
We study the K-complexity for system size $N=16$ with total spin $\mathcal{S}=2$ and parity $\mathcal{P}=+1$. We choose the initial operator at $i= 7$, i.e., $S_7^z + S_{10}^z$. We choose the integrability breaking parameter $H_d = S_{(N+1)/2}^z$ at the middle of the chain, and we study the non-integrable case for $\epsilon = 0.5$. For this system size and symmetry sector choice, the Krylov dimension is of the order $\mathcal{K} \sim 4000$, which is similar to the TFIM case studied earlier. \\
\newline
\textbf{Results:} In this example, we again find that the K-complexity shows behavior similar to the open TFIM results. The complexity grows initially, then decays followed by saturation. The saturation values are similar for integrable and non-integrable cases (Fig.\,\ref{fig:LanopenXXZ} (a)). Finally, as we increase the non-Hermitian coupling, the initial peak of the non-integrable case comes down closer to the integrable one (Fig.\,\ref{fig:LanopenXXZ} (b)).







\section{Appendix: A generalized version of spread complexity} \label{biL_spread}

Our approach can be directly implemented to compute spread complexity \cite{Balasubramanian:2022tpr}. For the spread complexity, one can follow a similar algorithm. The only difference is that in this case the evolution is of a state (unlike operator) and is generated by a non-Hermitian Hamiltonian (instead of Lindbladian). The job is then to tridiagonalize the Hamiltonian into the form \cite{Gruning}
\begin{align}
H = \begin{pmatrix} a_{0}&b_{1}&&&&0\\c_{1}&  a_{1}& b_{2}&&&\\&c_{2}&a_{2}&\ddots &&\\&&\ddots &\ddots &b_{m}&\\&&&c_{m}&a_{m}&
    \ddots\\0&&&&\ddots&\ddots\\\end{pmatrix}\,. \label{hamtri}
\end{align}
The basis is individually generated by $\ket{p_0}$ and $\ket{q_0}$, and they are bi-orthonormal according to \eqref{biortho}. Starting from by some initial state $\bra{w_0}$ and $\ket{u_0}$, one follows the following recursive algorithm \cite{Gruning}:
\begin{align}
    \ket{Q_{j+1}} = (H - a_j) \ket{q_j} - c_j \ket{q_{j-1}}\,,~~~~  
    \bra{P_{j+1}} = \bra{p_{j}} (H - a_j) - \bra{p_{j-1}} b_j\,,
\end{align}
where
\begin{align}
    a_j = \braket{p_j|H|q_j}\,, &~~~ b_{j+1} = ||Q_{j+1}||\,, ~~~ c_{j+1} = \frac{1}{b_{j+1}} \braket{P_{j+1}|Q_{j+1}}\,, \\
    \bra{p_{j+1}} &= \frac{1}{c_{j+1}} \bra{P_{j+1}}\,,~~~ \bra{q_{j+1}} = \frac{1}{b_{j+1}} \bra{Q_{j+1}}\,.
\end{align}
The algorithm recasts the Hamiltonian into the form \eqref{hamtri} (and thereby \eqref{LI}). Therefore, one can choose an initial state (in a lattice system, this can be thought of as a ground state of some arbitrary Hamiltonian, see \cite{Bhattacharjee:2022qjw} for an example) and evolve this with the non-Hermitian Hamiltonian. In the case of a Hermitian Hamiltonian, the tridiagonalized matrix form of the Hamiltonian is usually known to have real, but nonzero, diagonal coefficients \cite{Balasubramanian:2022tpr, Balasubramanian:2022dnj}. Hence, we expect that after the tridiagonalization of a non-Hermitian Hamiltonian using bi-Lanczos algorithm, the diagonals would have both real and imaginary parts. This was not encountered for the Lindbladian because the closed system tridiagonalized Liouvillian has all the vanishing diagonals. 
Real diagonals give rise to a phase factor in the solutions of the wavefunctions, whereas any imaginary part results a decay. Hence we expect the overall behavior of the spread complexity evolving under a non-Hermitian Hamiltonian to qualitatively mimic the operator K-complexity for open quantum systems. Some of these aspects will be covered in an upcoming work by one of the authors (A.B.) \cite{spreadupcoming}. 

\bibliographystyle{JHEP}
\bibliography{references}

\end{document}